\documentclass[
	aps, 
        pra,
        superscriptaddress,
        twocolumn,
	10pt,
	floatfix, 
        nobibnotes,
        nofootinbib,
	tightenlines,
        longbibliography
]{revtex4-2}

\usepackage[final]{graphicx}
\usepackage{times,bbm,amsmath,amssymb}
\usepackage{epsfig,color}
\usepackage{xcolor}
\usepackage{hyperref}
\hypersetup{
    colorlinks = true
}
\usepackage{cleveref}
\usepackage{microtype}

\usepackage{booktabs}
\usepackage{float}
\usepackage[caption = false]{subfig}

\usepackage[greek,english]{babel}
\usepackage{thumbpdf,enumerate}
\usepackage{booktabs}
\usepackage{sidecap}
\usepackage{pst-grad,bm}
\usepackage{epigraph}
\usepackage{longtable}
\usepackage{ulem}

\usepackage{acronym}
\usepackage{physics}
\usepackage{dsfont}
\usepackage{tabularx}
\usepackage{mathtools}
\usepackage{physics}

\usepackage{easyReview}
\usepackage{hyperref}
\usepackage{comment}

\begin{document}

\title{Variational approach to photonic quantum circuits via the parameter shift rule}

\author{Francesco Hoch}
\affiliation{Dipartimento di Fisica - Sapienza Universit\`{a} di Roma, P.le Aldo Moro 5, I-00185 Roma, Italy}

\author{Giovanni Rodari}
\affiliation{Dipartimento di Fisica - Sapienza Universit\`{a} di Roma, P.le Aldo Moro 5, I-00185 Roma, Italy}

\author{Taira Giordani}
\affiliation{Dipartimento di Fisica - Sapienza Universit\`{a} di Roma, P.le Aldo Moro 5, I-00185 Roma, Italy}

\author{Paul Perret}
\affiliation{DER de Physique - Université Paris-Saclay, ENS Paris-Saclay, 91190, Gif-sur-Yvette, France}

\author{Nicol\`{o} Spagnolo}
\affiliation{Dipartimento di Fisica - Sapienza Universit\`{a} di Roma, P.le Aldo Moro 5, I-00185 Roma, Italy}

\author{Gonzalo Carvacho}
\affiliation{Dipartimento di Fisica - Sapienza Universit\`{a} di Roma, P.le Aldo Moro 5, I-00185 Roma, Italy}

\author{Ciro Pentangelo}
\affiliation{Ephos, Viale Decumano 34, 20157 Milano, Italy}

\author{Simone Piacentini}
\affiliation{Istituto di Fotonica e Nanotecnologie, Consiglio Nazionale delle Ricerche (IFN-CNR), 
Piazza Leonardo da Vinci, 32, 20133 Milano, Italy}

\author{Andrea Crespi}
\affiliation{Istituto di Fotonica e Nanotecnologie, Consiglio Nazionale delle Ricerche (IFN-CNR), 
Piazza Leonardo da Vinci, 32, 20133 Milano, Italy}
\affiliation{Dipartimento di Fisica, Politecnico di Milano, Piazza Leonardo da Vinci 32, 20133 Milano, Italy}

\author{Francesco Ceccarelli}
\affiliation{Ephos, Viale Decumano 34, 20157 Milano, Italy}
\affiliation{Istituto di Fotonica e Nanotecnologie, Consiglio Nazionale delle Ricerche (IFN-CNR), 
Piazza Leonardo da Vinci, 32, 20133 Milano, Italy}

\author{Roberto Osellame}
\affiliation{Ephos, Viale Decumano 34, 20157 Milano, Italy}
\affiliation{Istituto di Fotonica e Nanotecnologie, Consiglio Nazionale delle Ricerche (IFN-CNR), 
Piazza Leonardo da Vinci, 32, 20133 Milano, Italy}

\author{Fabio Sciarrino}
\affiliation{Dipartimento di Fisica - Sapienza Universit\`{a} di Roma, P.le Aldo Moro 5, I-00185 Roma, Italy}

\begin{abstract}

{In the era of noisy intermediate-scale quantum computers, variational quantum algorithms are promising approaches for solving optimization tasks by training parameterized quantum circuits with the aid of classical routines informed by quantum measurements. In this context, photonic platforms based on reconfigurable integrated optics are an ideal candidate for implementing these algorithms. Among various techniques to train variational circuits, the parameter shift rule enables the exact calculation of cost-function derivatives efficiently, facilitating gradient descent-based optimization. In this paper, we derive a formulation of the parameter shift rule for computing derivatives and integrals tailored to reconfigurable optical linear circuits and based on the Boson Sampling paradigm. This allows us to naturally embed common types of experimental noise, such as partial distinguishability and mixedness of the states, thus obtaining a resilient approach. Finally, we employ the developed approach to experimentally test variational algorithms with single-photon states processed in a reconfigurable 6-mode universal integrated interferometer. Specifically, we apply the photonic parameter shift rules to the variational implementation, on a photonic platform, of both an eigensolver and a  Universal-Not gate.}

\end{abstract}

\maketitle

\section{Introduction}
\begin{figure}[t]
    \centering
    \includegraphics[width=\columnwidth]{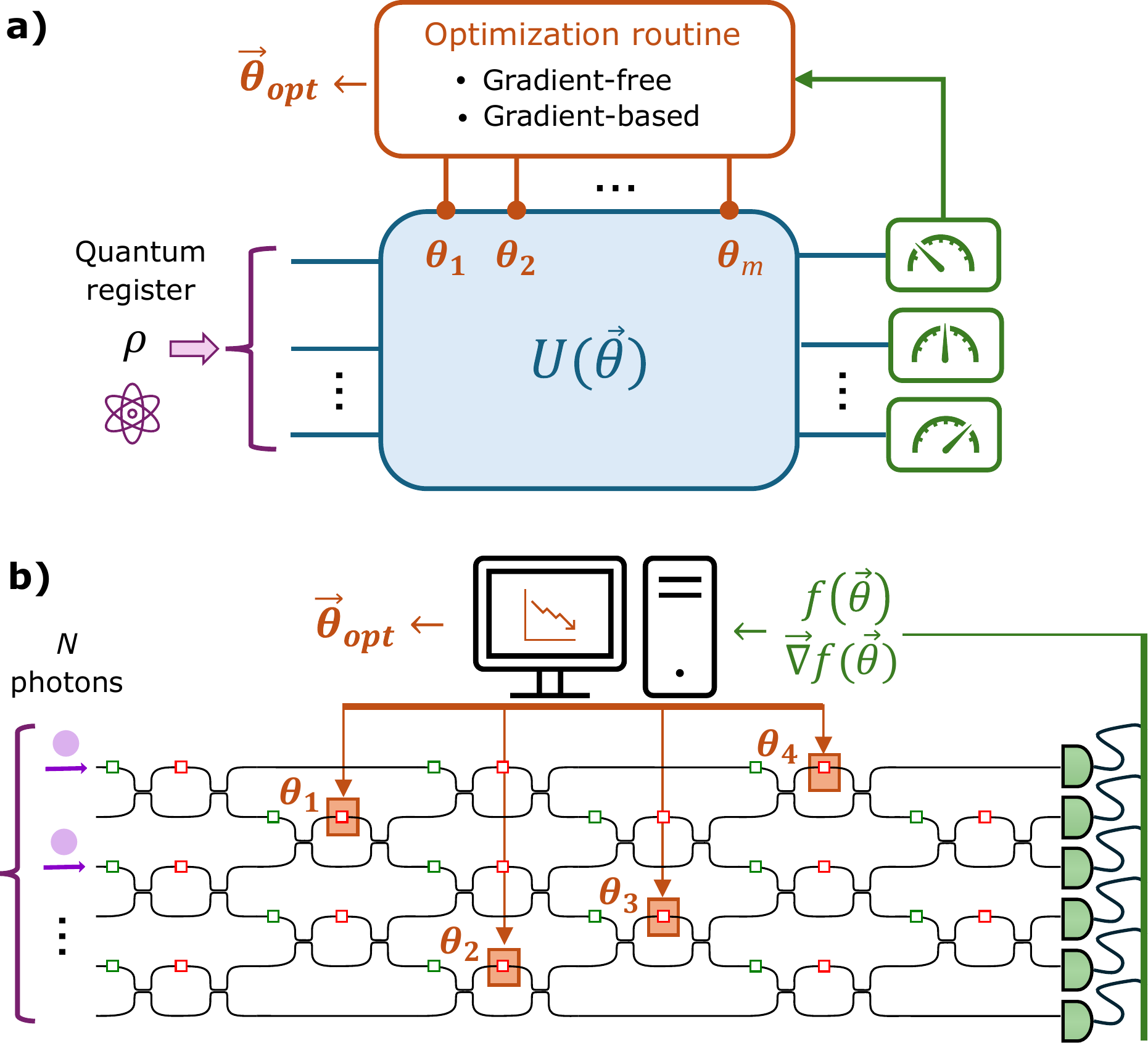}
    \caption{\textbf{Variational quantum circuits and the photonic parameter shift rule.}  a) A variational quantum circuit is a quantum device that takes in input a quantum register of qubits $\rho$ which evolves through a parametric unitary transformation $U(\Vec{\theta})$. The measurement outcomes on the output states fed a classical gradient-free or gradient-based optimization routine to change the parameters of the circuit and minimize a particular cost function.
    b) The photonic version replaces the variational circuit with a multi-port parametric interferometer in which $N$ photons evolve. 
    The output photon distribution depends on the internal phases of the optical circuit (green and red rectangles) through trigonometric functions. This makes it feasible for the circuit to estimate both the value and the gradient of such a class of functions via the parameter shift rule. In the figure, we display the 6-mode universal integrated interferometer used for testing the method in 1- and 2-photon experiments.}
    \label{fig:concept}
\end{figure}

In the current era of noisy quantum devices on an intermediate scale (NISQ), variational quantum algorithms (VQA) represent one of the main strategies to overcome some of the technical challenges implied by the use of noisy quantum processors. In fact, VQA opens up the possibility of applying current quantum devices to relevant applications \cite{Cerezo2021, Rev_Noisy_Var}. These algorithms are based on the application of a hybrid quantum-classical approach. In the context of VQAs, a particular quantum/classical problem is encoded in a cost function whose minima represent the desired solution of the problem which can be efficiently evaluated via measurements carried out at the output of a parametric quantum circuit. In such a way, a classical optimizer can be employed to iteratively tune the parameters of the quantum circuit in order to minimize the cost function and thus find the desired optimal solution (See Fig.~\ref{fig:concept}a).
This type of algorithms has been successfully applied to several different tasks, such as quantum dynamics simulation \cite{Endo2020,Yuan2019}, quantum chemistry \cite{Cao2019, Grimsley2019}, variational eigensolvers \cite{Peruzzo2014,Santagati2018, Maring2024},  quantum cloning machines \cite{hoch2024variational}, quantum metrology \cite{Cimini2024}, and optimization of quantum algorithms \cite{agresti2024, Zhu2019, Wang2024}. In the context of photon-based quantum computation, reconfigurable integrated circuits do represent one of the main platforms which can be used to implement and test VQAs, as they allow for the realisation of relatively complex linear optical interferometers with a large number of adaptive parameters while maintaining a high degree of stability and compactness \cite{Wang2019, Pelucchi2021, Giordani2023}.

One of the principal challenges of the VQA is the application of the classical optimizer to the quantum circuit. Indeed, many classical optimisation algorithms are based on the estimation of the cost function gradient through the use of the finite difference method, in which derivatives are replaced by differences between function evaluations at nearby points, that are distant by a small amount $\varepsilon \ll 1$.
While for numerical algorithms fully running on classical hardware this is a feasible approach, such a task becomes impractical in the case of noisy quantum devices. Indeed, in the latter case it is necessary to find a value of $\varepsilon$ small enough to provide a good approximation of the derivative, while being at the same time large enough so that the signal-to-noise ratio of the derivative estimation is small.
One of the possible solutions to solve such an intrinsic problem of derivatives calculations is given by the use of gradient-free optimization algorithms. Here only the comparison of the function value at relatively distant points and a heuristic on how to search the minimum are required. Many experimental implementations of VQAs in photonic platforms which can be found in the literature have relied on such gradient-free approaches \cite{Santagati2018, Paesani2017, hoch2024variational, Jaek2019, Maring2024}.
In the circuit-based quantum computing paradigm a different solution has been developed, known as \textit{parameter shift rule} \cite{schuld2019evaluating, mitarai2018quantum, Wierichs2022}.
This method provides an unbiased estimation of the derivatives of a function, i.e. without approximations, by evaluating the function on a set of distant points and thus enabling the evaluation of derivatives on noisy quantum hardware. 
This is possible since it can be shown that the output probabilities of a quantum circuit are trigonometric functions of the variational parameters, thus allowing the calculation of derivatives by evaluating the functions at points distant multiples of $\pi$.
While the parameter shift rule has already been developed in a circuit-based approach, its translation to a photonic scenario poses some challenges, and is in principle associated to additional overheads in terms of number of ancillary modes and photons. This motivates the need to derive a general parameter shift rule that is conceived to operate natively for photonic systems.
A first approach to the parameter shift rule for photonic architectures can be found in Ref. \cite{Cimini2024} and a similar result were found in \cite{Gan2022} applied to the expressiveness of photonic circuits for machine learning applications.

In this paper, we derive a general photonic version of the parameter shift rule applied to linear optical circuits using the Boson Sampling formalism \cite{Aaronson2013}, that allows us to incorporate the typical sources of noise of a multi-photon experiment into the derivation.
We experimentally test for the first time this approach in a universal integrated optical chip. The photonic parameter shift rule is employed for the calculation of gradients and integrals in order to apply gradient-based optimization algorithms directly on the quantum hardware.
In particular, we apply such a strategy to a classical minimization problem known as variational eigensolver \cite{Sakurai1993}. Analogously, the same method is applied to a notorious problem in quantum information known as the optimal  Universal-Not gate \cite{Buzcircek2000, DeMartini2002, Ricci2004}.

The structure of the work is the following. In Sec II, the theory behind the parameter shift is presented by deriving the analogous version for linear photonic circuits.
Specifically, it is demonstrated that the probability of an event at an interferometer's output as its phases change takes the shape of a trigonometric series, the degree of which is proportional to the number of photons in the circuit. This makes it possible to derive parameter shift rules analogous to those developed for quantum circuits.
It is also shown that the main sources of noise in photonic experiments, such as partial photon distinguishability and losses, do not affect the functional form and thus the application of the parameter shift rule remains robust. In addition, it is shown how the theory behind the parameter shift rule can also be extended to the calculation of integrals over some of the parameters of the circuit.
In Sec. III, we experimentally test the developed theory using $1$ and $2$ photons in a $6$-mode reconfigurable photonic circuit. First, a 3-mode interferometer is used to show how the parameter shift rule changes as the structure of the interferometer and the number of photons used vary. Subsequently, applications to compute derivatives and integrals are used to solve two variational problems, specifically a variational eigensolver and a variational Universal-Not gate. Finally, in Sec. IV, we present a summary of the results covered and an overview of the future prospects.


\section{Parameter shift rule for linear photonic circuit}
For the computation of the gradient of a cost function defined through a parametric quantum device one of the most established ideas is to use the so-called parameter shift rule \cite{Wierichs2022}, whose purpose is to exactly and efficiently measure the derivative of a function by evaluating the function itself at shifted points "far" from each other.

The idea behind the method is that for a large variety of quantum circuits, the dependence of the measurement output probabilities on the variational parameters can be expressed as a finite Fourier series.
In particular, for most cases, the expression can be further simplified by noting that the angular frequencies are integer, meaning that in general one can write the output probabilities as:
\begin{equation}
    f(x) = a_0 + \sum_{l = 1}^R a_l \cos(l x)+ b_l \sin(l x)
    \label{eq:trig_func_s}
\end{equation}
Due to this functional form, we can reconstruct the function by evaluating it in $2R+1$ points. This can be related to the trigonometric interpolation problem \cite{trigonometric2015}. 

Moreover, its derivative $f'(x_0)$ in a given point $x_0$ can be evaluated by following the approach of Ref. \cite{Wierichs2022}.

\textbf{Parameter shift rule:} \textit{Ghiven a function $f(x)$ in the form of Eq.~\ref{eq:trig_func_s} the derivative of the function in a point $x_0$ can be computed by evaluating the function in a set of $2R$ translated points $x_0+x_k$ with $\{x_k = \frac{2k+1}{2R}\pi | k \in \{0,\dots, 2R-1\}\}$, using the formula:}
\begin{equation}
    \dv{f}{x}\biggl|_{x = x_0} = \sum_{k=0}^{2R-1}f\left(x_0+\frac{2k+1}{2R}\pi\right) \frac{(-1)^{k-1}}{4R\sin^2(\frac{2k+1}{4R}\pi)}
    \label{eq:parshift}
\end{equation}
For more specific details, we refer to the specialized literature \cite{Wierichs2022}.

\subsection{Photonic phase}

Now we show how the parameter shift rule behaves in the photonic framework, specifically when considering the linear optical model comprising single-photon resources interacting in a multi-mode interferometer composed of linear optical elements \cite{kok2007linear}.
We consider $N$ indistinguishable photons at the input of a linear $m$-mode interferometer described by the unitary matrix $U$, which defines the transformation between input and output mode operators as $b^\dag_j = \sum_{k} U_{jk} a^\dag_k$. Such unitary matrix $U$ admits in general a decomposition in terms of a set of single-mode phase shifts and balanced two-mode beam splitters, as described in Ref. \cite{Clements2016}. Thus, any output probability will in general be a function of a set of single-mode phase shifts $\varphi$.

Within the linear optical framework, the input distribution of the photons is defined by the mode occupation list $\Vec{r} = (r_1, \dots r_m)$ where $r_j$ photons are in the input mode $j$. Similarly, we can define the set of output states as  $\Vec{s} = (s_1, \dots s_m)$.
For convenience, we define the mode assignment list \cite{Tichy2012} as:
\begin{equation}
    \Vec{d}(\Vec{r}) = (\underbrace{1,\dots,1}_{r_1}, \underbrace{2,\dots,2}_{r_2}, \dots, \underbrace{m,\dots,m}_{r_m})
\end{equation}
We also define the normalization constant $\mu(\Vec{r}) = \prod_{j = 1}^m r_j !$.
Finally, we define the scattering matrix $M$ associated with the input and output states $\Vec{r}$ and $\Vec{s}$ as the $N\times N$ matrix defined as
\begin{equation}
    M_{hj} = U_{\Vec{d}(\Vec{s})_h, \Vec{d}(\Vec{r})_j}
\end{equation}
The probability to obtain the output state $\Vec{s}$ from the input state $\Vec{r}$ can be computed as \cite{brod2019photonic}:
\begin{equation}
    P(\Vec{s}|\Vec{r}) = \frac{\abs{\text{Per}(M)}^2}{\mu(\Vec{r})\mu(\Vec{s})}
    \label{eq:ptrans}
\end{equation}
where $\text{Per}(M)$ is the permanent of the scattering matrix $M$.

We now discuss the scenario in which the unitary evolution $U(\varphi)$ of the interferometer depends on a single phase shift $\varphi$, and we prove that the probability function of Eq.~\eqref{eq:ptrans} can be written in the functional form of Eq.~\eqref{eq:trig_func_s}.

In general the unitary transformation $U(\varphi)$ can be divided into three parts;
\begin{equation}
    U(\varphi) = U^{(2)} \Phi^{(r)} U^{(1)}
\end{equation}
where $U^{(1)}$ and $U^{(2)}$ are unitary matrices describing the linear interferometer before and after the phase under exam and $\Phi^{(r)}$ is the unitary matrix of the phase $\varphi$ in the mode $r$ that can be expressed as $\Phi^{(r)}_{hj} = \delta_{hj} e^{i\delta_{hr}\varphi}$, where $\delta_{hj}$ is the Kronecker delta.
Through this decomposition, it can be easily shown that each element of the unitary transformation is in the functional form:
\begin{equation}
    U_{hj}(\varphi) = \sum_{k} U^{(2)}_{hk} U^{(1)}_{kj} e^{i\delta_{kr}\varphi} \coloneqq c_{hj} + d_{hj} e^{i \varphi}
    \label{eq:unitary_elements}
\end{equation}
Using the definition of the permanent, we can expand the formula for the probability shown in Eq. \eqref{eq:ptrans} as:
\begin{equation}
    P(\Vec{s}\,|\Vec{r}) = \frac{1}{\mu(\Vec{r})\mu(\Vec{s})} \biggl|\sum_{\sigma \in S_N} \prod_{k = 1}^N U_{\sigma(\Vec{d}(\Vec{s})_k), \Vec{d}(\Vec{r})_k}\biggr|^2
\end{equation}
where $S_N$ is the set of permutations of $N$ elements.
Given the functional form in Eq.~\eqref{eq:unitary_elements} of the coefficient of the unitary matrix, and since the product of $N$ binomial terms of degree $1$ is a polynomial with degree $N$, we can write
\begin{equation}
    \sum_{\sigma \in S_N} \prod_{k = 1}^ N U_{\sigma(\Vec{d}(\Vec{s})_k), \Vec{d}(\Vec{r})_k} = \sum_{k=0}^N c_k e^{ik\varphi}
    \label{eq:unitary_series}
\end{equation}
This gives us the final result
\begin{equation}
    P(\Vec{s}\,|\Vec{r}) = \frac{1}{\mu(\Vec{r})\mu(\Vec{s})} \sum_{k=0}^N d_k e^{ik\varphi}+d^*_k e^{-ik\varphi}
    \label{eq:unitary_expansion}
\end{equation}
where $d_k = \sum_{h = 0}^{N-h} c^*_h c_{h+k}$. This equality can be rewritten as:
\begin{equation}
   P(\Vec{s}\,|\Vec{r}) = \sum_{k=0}^N a_k cos(k\varphi) + b_k sin(k \varphi)
   \label{eq:pshift_phot}
\end{equation}
proving that the output probabilities from a linear interferometer follow the functional form of Eq.~\eqref{eq:trig_func_s}  with $R = N$, i.e. equal to the number of photons at the input of the interferometer.

When the variational parameters are associated with more than one phase, the result is similar. The functional form is a multi-parameter Fourier series of degree $N$ in each of the phase variables.
The demonstration is identical to the single-parameter case. 
For example, in the case of two phases $\varphi_1$ and $\varphi_2$, the output probability will be a Fourier series in the first phase $\varphi_1$ as in Eq.~\eqref{eq:pshift_phot} whose coefficients $a_k(\varphi_2)$ and $b_k(\varphi_2)$ are themselves Fourier series in the second phase $\varphi_2$.

\subsection{Role of experimental noise}
Notably, if one relaxes the assumptions of full indistinguishability of the photonic resources and considers a generic input state that includes mixed states and partial distinguishability in the photons' internal degrees of freedom,
the functional form shown in Eq.\eqref{eq:pshift_phot} does not change. To prove this, we employ the formalism presented in Ref. \cite{Shchesnovich2015}. There, it is shown that the transition probabilities can be computed as:
\begin{multline}
    P(\Vec{s}\,|\Vec{r}) = \frac{1}{\mu(\Vec{r})\mu(\Vec{s})} \sum_{\sigma_1, \in S_N} \sum_{\sigma_2 \in S_N} J(\sigma_1, \sigma_2) \times \\
    \times \prod_{k = 1}^N U^*_{\sigma_1(\Vec{d}(\Vec{s})_k), \Vec{d}(\Vec{r})_k} \prod_{h = 1}^N U_{\sigma_2(\Vec{d}(\Vec{s})_h), \Vec{d}(\Vec{r})_h} 
\end{multline}
where $J(\sigma_1, \sigma_2)$ is the spatial distinguishability matrix, which is a hermitian positive semi-definite matrix indexed by two permutations $\sigma_1$ and $\sigma_2$ that takes into account not only the partial distinguishability of the photon but also the photon states mixedness, the sensitivities of the output detectors and, more generally, photon losses.

Using the same reasoning as in Eq.~\eqref{eq:unitary_series}, we can write:
\begin{equation}
    \prod_{h = 1}^N U_{\sigma(\Vec{d}(\Vec{s})_h), \Vec{d}(\Vec{r})_h}  = \sum_{k=0}^N c_k^{(\sigma)} e^{ik\varphi}
\end{equation}
Using the fact that $J$ is a Hermitian matrix, one can show that Eq.~\eqref{eq:unitary_expansion} with coefficients $d_k$ will be now given as:
\begin{equation}
    d_k = \sum_{h = 0}^{N-k} \sum_{\sigma_1, \in S_N} \sum_{\sigma_2 \in S_N} J(\sigma_1, \sigma_2) c^{*(\sigma_1)}_h c^{(\sigma_2)}_{h+k}
\end{equation}
Thus, analytically, the transition probabilities at the output of a linear interferometer can always be expressed in terms of a Fourier expansion as in Eq.\eqref{eq:trig_func_s}, irrespectively of the indistinguishability properties of the input photons.

Furthermore, we note that the value of the parameter $R$ of Eq.~\eqref{eq:trig_func_s} in the photonic picture is not strictly related to the number of photons at the input of the interferometer itself but to the maximum number of photons that can interact with a considered phase, given the input-output Fock state configuration. This, in turn, makes it possible to decrease the number of measurements required depending on the structure of the interferometer and the input and output configurations considered. A complete proof of this behaviour is reported in {Appendix A}.

\subsection{Beyond derivatives computation}
Until now we have described a method to compute the derivatives of a function with the parameter shift rule. However, the knowledge of the functional form of Eq.~\eqref{eq:trig_func_s} allows to extend this method to different kinds of computations.
As an example, some cost functions in variational algorithms scenarios involve the average over a set of states that can be encoded in one or more parameters of the variational circuit \cite{coyle2022progress}.
This problem can be translated into the calculation of the integral of the function $f(x)$ weighted with a periodic distribution $g(x)$
\begin{equation}
    {M} = \int_{-\pi}^{\pi} \dd x f(x) g(x) 
\end{equation}
if we define the parameters
\begin{equation}
    c_k = \int_{-\pi}^{\pi} \dd x \; g(x) \frac{\sin(\frac{2R+1}{2}x)}{\sin(\frac{x}{2}-\frac{k}{2R+1}\pi)}
\end{equation}
then the integral can be rewritten as
\begin{equation}
    {M} = \sum_{k= 0}^{2R} \frac{(-1)^k c_k}{2R+1} f\left(\frac{2k}{2R+1}\pi\right) 
\end{equation}
In general, this requires $2R+1$ evaluations of the function $f(x)$. However, if additional information on the parity of the weight distribution $g(x)$ is available, then one can reduce the number of measurements to $2R$. Indeed, as an example, if we want to compute the average of the function $\langle f(x) \rangle$ (i.e. $g(x) = \frac{1}{2\pi}$) we can use the following equation
\begin{equation}
    \langle f(x) \rangle = \frac{1}{2R} \sum_{k=0}^{2R-1} f\left(\frac{k\pi}{R}\right)
\end{equation}
As a relevant aspect, we observe that there are no approximations in the derivation. This implies that the error is only due to the evaluation of the function $f(x)$ and not to the use of a finite sample for the evaluation of the mean.

The formal derivation of the results given in this section and the generalisation of the formulae for a generic distribution $g(x)$ are given in Appendix B.

\begin{figure*} [t]
    \centering
    \includegraphics[width=\textwidth]{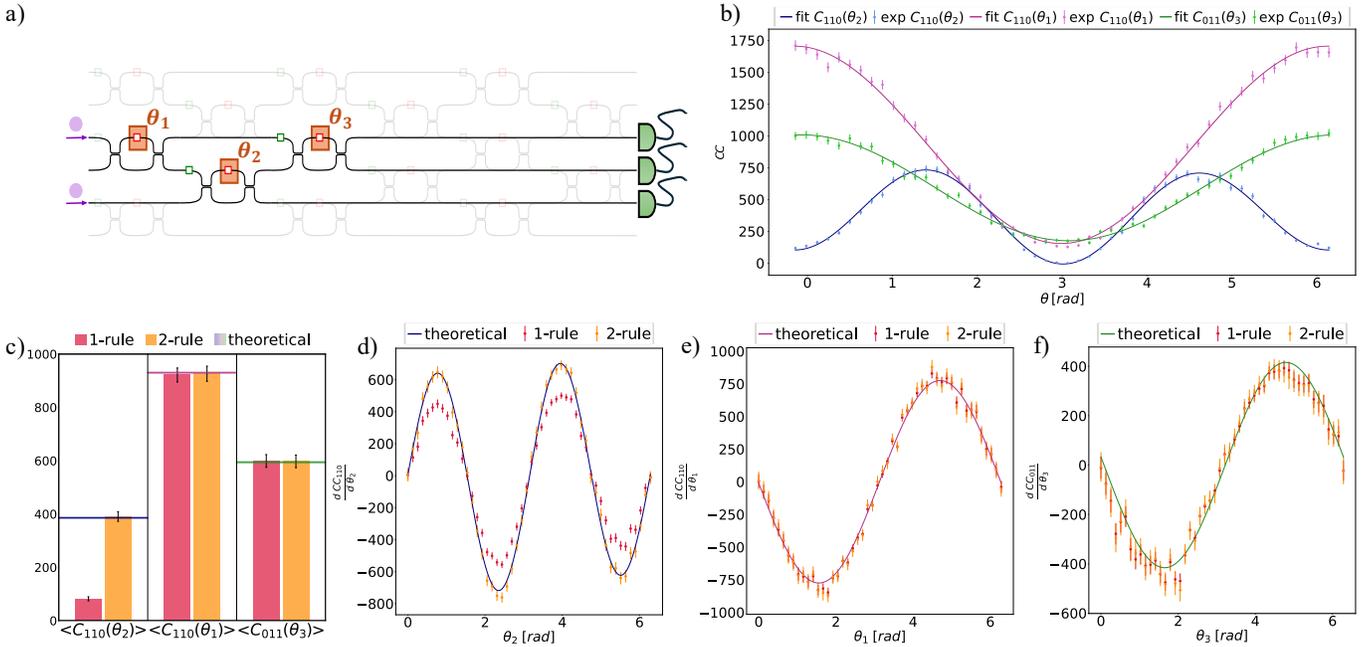}
    \caption{\textbf{Parameter shift rule in a 3-mode interferometer.} a) universal 3-mode interferometer inside the 6-mode one. In orange, are the variational phases where the parameter shift rule is tested. 
    b) Two-photon counts for $C_{110}(\theta_2)$ (blue), $C_{110}(\theta_1)$ (purple) and $C_{011}(\theta_3)$ (green) by varying the respective angle in $[0,\, 2\pi]$.
    c) 1- and 2-photon rules applied for estimating the averages of the two-fold coincidences $\langle C_{110}(\theta_2)\rangle$, $\langle C_{110}(\theta_1)\rangle$ and $\langle C_{011}(\theta_3)\rangle$. In red the experimental estimates of the 1-photon rule, in orange the 2-photon rule results and in blue the theoretical expectation.  In d) the partial derivative with respect to $\theta_2$ computed via the 1-photon rule (red) and the 2-photon rule (orange).
    The solid blue line in b) is the fit of the data and the one in d) is its partial derivative. e-f) Partial derivatives $\frac{\partial{C_{110}}}{\partial{\theta_1}}$ and $\frac{\partial{C_{011}}}{\partial{\theta_3}}$ computed via the two rules parameters. In both cases, the two estimates are compatible within the experimental uncertainties due to the Poisson statistics of single-photon counts. The solid lines are the theoretical expectation for the partial derivative.}
    \label{fig:tritter}
\end{figure*}

\section{Experimental implementation}

In this section, we will show how the derivative estimation rules described in the previous section can be applied to an actual implementation of a variational circuit in a photonic platform. Specifically, we tested the effectiveness of the parameter shift rule in a quantum photonic experiment by exploiting pairs of indistinguishable single-photon states and a 6-mode programmable universal integrated photonic circuit.
Before going into the details of the results, we briefly describe the experimental platform. 

The single-photon source is a parametric down-conversion based on a Beta-Barium-Borate nonlinear crystal.
The source generates photon pairs at $\lambda = 785\, \text{nm}$. Spectral filtering, time-synchronization, and polarization control over the two-photon states allowed us to reach a level of indistinguishability up to $0.99 \pm 0.01$. The two photons are coupled through single-mode fibers to a 6-mode integrated optical circuit. The architecture of the device follows the universal scheme of Ref. \cite{Clements2016} pictured in Fig.\ref{fig:concept}b. Each red and green square represents a tunable phase of the interferometer while the crossing stands for unbiased beam-splitters. The interferometer can be programmed by changing the internal phase values to perform any unitary transformation over the optical modes. The chip is fabricated through the femtosecond-laser-writing technology \cite{Corrielli2021}. A set of 30 heaters that induce local changes in the refractive index of the waveguides by the thermal-optic effect enable the full reconfigurability of the internal phases. Further details on the performance and specs of such an integrated device can be found in Refs. \cite{Pentangelo2024,Giordani24, hoch2024variational}.
Finally, a multimode fiber array collects the two photons in the outputs and sends the signal to avalanche photo-diode single-photon detectors. A time-to-digital converter device processes two-fold coincidences among the detectors.

The whole platform is controlled by a classical routine fed by the function values and gradient, computed by the parameter shift rule, both estimated from the photon counts.
Then, the variational algorithm works as follows. In order to compute the gradient of a cost function which is related to the output parameters of the circuits, the rule presented in Eq.\eqref{eq:parshift} requires shifting the variational phases of the interferometer (in orange in Fig. \ref{fig:concept}b) of a given amount. 
To this end, the phases of the chip under analysis can be suitably programmed to perform such a task, with the collection of single-photon counts and the output two-photon distribution giving an estimate of both, the function and of its partial derivatives. 
Then, a gradient-based optimization performed by a classical routine iteratively updates the on-circuit variational parameters and predicts their optimal values for the problem under investigation.

\subsection{Testing one- and two-photon rules}

This preliminary section aims to provide insight into the operation and effectiveness of a $N$-photon parameter shift rule that follows from Eq. \eqref{eq:parshift} with $R=N$. In particular, we focus on the verification of 1- and 2-photon rules by sending two indistinguishable photons from the source to the integrated interferometer. We select a portion of the 6-mode circuit that implements a 3-mode transformation and we inject the photons in the input configurations $[1,0,1]$, where the string expresses that the two photons are injected from modes 1 and 3 (see Fig. \ref{fig:tritter}a). 

The structure of the interferometer is the ideal testbed to understand how the $N$-photon rule works. The transition function that links the input string to the photon counts in one of the outputs $\{[1,1,0], \, [1,0,1], \, [0,1,1] \}$ is in the form of Eq.~\eqref{eq:trig_func_s} and depends in general on the three phases highlighted in orange in Fig. \ref{fig:tritter}a), so that in general $f=f(\theta_1, \theta_2, \theta_3)$. In particular, we consider the functions $C_{110}(\theta_2)$, $C_{110}(\theta_1)$ and $C_{011}(\theta_3)$ where $C_{ijk}(\theta)$ correspond to the two-photon coincidence counts for two photons exiting from the output configuration $[i,j,k]$, while the angles in the parentheses are the phases varied in the application of the parameter shift rule (see Fig. \ref{fig:tritter}b). We expect from the previous derivation that the $n$-photon rule with $N$ equal to the number of photons in the circuit, in our case $N=2$, always works in order to estimate both the function averages and partial derivatives. However, the structure of the circuit and the presence of possible symmetries could allow to use a simplified photon rule that involves a lower number of photons.
We discuss in detail such an intuition.

In Fig.~\ref{fig:tritter}c we test the 2-photon and the 1-photon rules 
for the estimation of the three averages $\langle C_{110}(\theta_2)\rangle$, $\langle C_{110}(\theta_1)\rangle$ and $\langle C_{011}(\theta_3)\rangle$. The estimation of $\langle C_{110}(\theta_2)\rangle$ is one of the cases that require the use of the two-photon rule for a correct computation. In fact, the estimation via the 1-photon rule does not provide a correct estimate of the average.
In contrast, if one looks at the phase $\theta_1$ in the photonic circuit, it is evident that only one of the two photons interacts with such a phase. This means that in principle the 1-photon rule could be enough for the estimation of $\langle C_{110}(\theta_1)\rangle$. To prove this, we compare the average of this function computed through the two rules.
Indeed, with respect to $\langle C_{110}(\theta_1)\rangle$, the 1-photon rule's estimate
is compatible with the 2-photon rules one within the experimental errors. Furthermore, both estimations are in accordance with the expected theoretical values.
Such a result confirms the effectiveness of the simplified rule that involves only one photon for some parameters and functions. 
A similar reasoning holds for the third average $\langle C_{011}(\theta_3)\rangle$. Indeed, this is another configuration in which only one photon is sensitive to changes in the parameter. Specifically, while the input state would allow in principle the interaction of two photons with the parameter $\theta_3$, the transition probability that links the input to the output $[0,\,1,\,1]$ 
has only one photon interacting with $\theta_3$. This allows us to use the 1-photon rule.
The rightmost panel of Fig. \ref{fig:tritter}c shows the estimates of $\langle C_{011}(\theta_3)\rangle$ with the two rules and their accordance with the expectation. 

We now move to the verification of the photonic parameter shift rule to compute the partial derivatives of a function $f=f(\theta_1, \theta_2, \theta_3)$. Specifically, here we are interested in the computation of $\frac{\partial{C_{110}}}{\partial{\theta_2}}$, $\frac{\partial{C_{110}}}{\partial{\theta_1}}$ and $\frac{\partial{C_{011}}}{\partial{\theta_3}}$.
As a first test, we report in Fig. \ref{fig:tritter}b the two-photon counts $C_{110}(\theta_2)$ collected by varying the parameter $\theta_2$ in $[0,\, 2\pi]$ and keeping constants the others. In Fig.\ref{fig:tritter}d, we show the partial derivative according to the 1- and 2-photon rules. We retrieved a similar behaviour of the one probed in the previous investigation. Only by employing the 2-photon rule, one can compute a correct estimate for the derivative, in accordance with the theoretical curve obtained by computing the analytical derivative of the data fit shown in Fig.\ref{fig:tritter}b. Finally, the experimental estimates $\frac{\partial{C_{110}}}{\partial{\theta_1}}$ and $\frac{\partial{C_{011}}}{\partial{\theta_3}}$ via the 2- and 1-photon rules are also shown. Again, for these configurations the 1-photon rule is sufficient for obtaining the correct computation of the partial derivatives, whose two estimates appear to be in accordance with the theoretical derivatives.

\begin{figure*}[t]
    \centering
    \includegraphics[width=\textwidth]{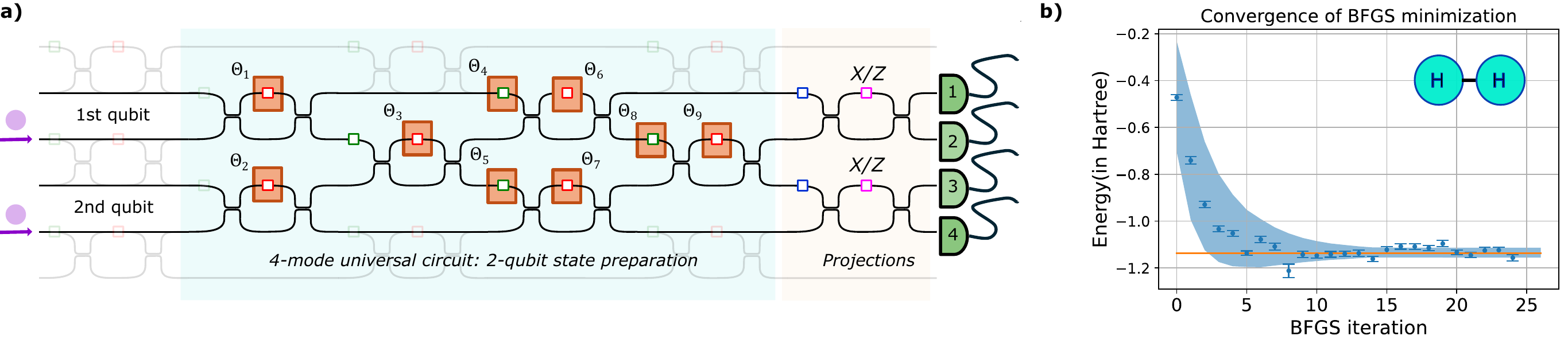}
    \caption{\textbf{Photonic variational eigensolver.} a) Structure of the variational circuit for evaluating the ground state energy of H2. Two dual-rail qubits are encoded in the two photons injected in modes 3 and 5 of the 6-mode interferometer. Then, we have a universal 4-mode interferometer that allows us to generate the ground state of the Hamiltonian (blue part of the circuit). The orange rectangles highlight the 9 variational phases. The last layer in yellow encodes the measurements of the operators $X$ and $Z$ on each qubit. b) Trend of the ground state energy value during the minimization by the BFGS method. The algorithm is fed by the gradient calculated by the quantum variational circuit via the 2-photon parameter shift rule. }
    \label{eigensolver}
\end{figure*}

Such a first experimental investigation demonstrates that the quantum variational circuit can be controlled suitably in order to calculate trigonometric functions and their partial derivatives, directly on device. Furthermore, it shows that the symmetries and structure of a given circuit can simplify the calculation by enabling the estimation of a function gradient via a lower-order parameter shift rule. This property could become significant in large-scale experimental implementations, since it would reduce the number of necessary measurements for a given function integrals and gradient estimations.

\subsection{Variational eigensolver}

After proving experimentally that the parameter shift rule, applied to a photonic testbed, correctly predicts derivatives of transition functions at the output of a linear optical interferometer, the natural next step is to variationally solve a practical problem by computing the gradients of the relevant cost function fully on-device.
Recently, such a problem of computing gradients with linear optical circuits via the parameter shift rule has been investigated to directly compute the so-called Fisher information in the context of quantum metrology 
\cite{Cimini2024}.
In our case, we employ the knowledge of the gradient of a function to implement a variational approximation protocol in order to compute the ground state energy of a hydrogen molecule ($\text{H}_2$).

Specifically, recall that the molecule bond energy depends on the nuclear separation between the two atoms. To obtain the expected minimal energy for each separation length, the \textit{OpenFermion} package provided the data and calculations \cite{McClean2020}. The minimum ground state energy is found at the known bond length of an $\text{H}_2$ molecule: $r = 0.7414 \,\text{\AA}$. 
Then, we can compute the Hamiltonian of an $\text{H}_2$ molecule at such a length
using \textit{OpenFermion} which gives the Hamiltonian in the form of a fermionic operator. The binary code function can then transform this operator into a qubits operators with lower qubit requirements, i.e. using two qubits instead of four. 
Thus, the Hamiltonian can be described by a sum of tensor products of single-qubit Pauli operators:
\begin{equation}
\label{hqubit}
    H_{qubit} = \alpha II +\beta ZI +\gamma IZ +\delta ZZ +\mu XX
\end{equation}
with coefficients that depend on the bond length. The values of the coefficients for an atomic separation equal to $r = 0.7414 \text{\AA}$ are reported in Table \ref{tab: ham_coeff}.
\begin{table}[htb]
\begin{center}
\begin{tabularx}{0.9\columnwidth} { 
   >{\centering\arraybackslash}X 
   >{\centering\arraybackslash}X 
   >{\centering\arraybackslash}X 
   >{\centering\arraybackslash}X
   >{\centering\arraybackslash}X }
   \toprule
    $\alpha$ & $\beta$ & $\gamma$ & $\delta$ & $\mu$ \\ 
    \midrule
    -0.340 & 0.394 & 0.394 & 0.011 & -0.181\\
\bottomrule
\end{tabularx}
\end{center}
\caption{\textbf{Coefficient of the hydrogen molecule Hamiltonian.} Parameters of the two-qubit Hamiltonian operator for a nuclear separation of $r = 0.7414 \,\text{\AA}$ between the two atoms. All the values are reported in Hartree energy units (Ha).}
\label{tab: ham_coeff}
\end{table}

The ground state of the molecule has an energy $E_0$ such that for any other state $E_0 \leq \langle \Psi |H|\Psi \rangle$. To find the ground state, one can therefore minimize $\langle \Psi |H|\Psi \rangle$ by making $\Psi(\theta_i)$ vary. 

Previous variational approaches to similar problems via photonic integrated circuits \cite{Peruzzo2014, Maring2024} employed a gradient-free classical optimization routine, such as the Nelder-Mead optimization \cite{Nelder1965}, with solely the expectation value of the energy computed by the quantum hardware. In particular, these experiments considered two dual-rail encoded qubits, single-qubit rotations, and a probabilistic entangling photonic CNOT gate to implement the variational searching of the ground state energy of a molecular compound.
Here we apply the introduced photonic parameter shift rule and perform the variational optimization via a gradient-based algorithm, with functional gradients directly computed on-device. Specifically, we encoded the variational eigensolver in a general 4-mode interferometer and used the circuit depth of the whole 6-mode device.

In our experiment, two indistinguishable photons enter the input 3 and 5 of the 6-mode chip and undergo a 4-mode unitary transformation (blue area in Fig.\ref{eigensolver}a) which gives $\ket{\Psi}$. In fact, each photon encodes a dual rail qubit in which the logical states correspond to the presence of the photon in one of the two paths. The blue area of the circuit can implement any 4-mode transformation by properly programming the 9 internal phases highlighted in orange. Then $\langle \Psi |H|\Psi \rangle$ is computed by measuring the expectation values associated to the Pauli operators of each term of the Eq.~\eqref{hqubit} (see Appendix D for more details). Before measuring the XX product, we change the base from ZZ to XX using a tunable Mach-Zehnder interferometer on each qubit (yellow region in Fig. \ref{eigensolver}a). All the other products share the same basis, and are thus measured at once, so each computation of $\langle \Psi |H|\Psi \rangle$ takes two measurements (see Appendix D).
For the minimization, we use the Broyden–Fletcher–Goldfarb–Shanno (BFGS) gradient-based optimization method \cite{Fletcher2000} on $\langle \Psi |H|\Psi \rangle$, with 9 variational phases represented by orange rectangles Fig.\ref{eigensolver}a and the gradient given by the 2-photons rule. The gradient computed by the quantum circuit at each iteration enters into the standard \textit{minimize} function of the python \textit{scipy} package \cite{2020SciPy-NMeth}. 

In Fig.~\ref{eigensolver}b we report the results of the minimization process.
Here, the dots represent the values of the energy during the optimization process as a function of the number of iterations of the minimization protocol. The final result is $E_{exp} = -1.150 \pm 0.012~\text{Ha}$, compatible with the theoretical value of $E_{th} = -1.137~\text{Ha}$ (orange line). Moreover, for comparison, we simulate the experiment 1000 times, with comparable finite statistics, and we plot a standard deviation around the average (blue region). Via this experiment, we show the possibility of applying the parameter shift rule in a photonic quantum computation paradigm, in order to compute the gradients of a suitably defined cost function computed by the evaluation of probabilities at the output of a linear optical interferometer. This, in turn, enables the possibility of using an efficient gradient-based optimization algorithm in order to variationally optimize such a cost function, thus unlocking the capability of running a large array of variational algorithms fully on-device.

\subsection{Variational implementation of a Universal-not gate}
\label{sec:U-not}

\begin{figure*}[t]
    \centering
    \includegraphics[width=\textwidth]{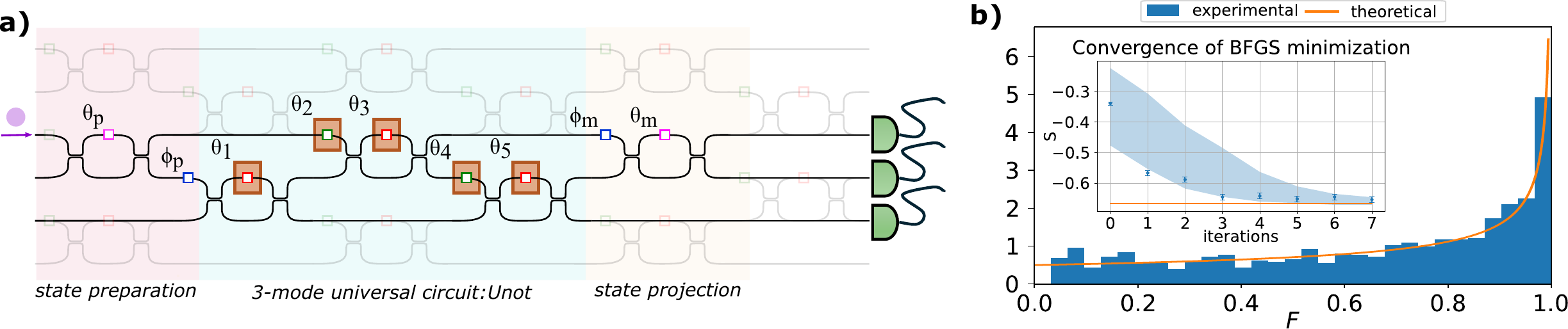}
    \caption{\textbf{Photonic universal NOT.} a) Structure of the variational circuit for the implementation of an optimal Universal-Not gate. A dual-rail qubit is prepared by injecting a photon in a tunable beam splitter controlled by the phases $(\theta_p, \phi_p)$ (red part of the circuit). After that, the state is injected in a three-mode universal interferometer implementing the variational Universal-Not gate controlled by the five phases $\theta_1, \dots, \theta_5$ (blue region). Finally, after post-selecting on the condition that no photon came out from the third photon, the state is projected on the orthogonal state $(\theta_m = \pi-\theta_p, \phi_m = \pi-\phi_p)$ and the fidelity is estimated by the resulting measurement. b) Distribution of the fidelity of the states used for the validation process, in orange the theoretical distribution. In the inset, we present the trend of the cost function during the minimization process with in orange the theoretical minimum.}
    \label{fig:Universal-not}
\end{figure*}

Finally, we present an application where the parameter shift rule is applied to compute the cost function as an average over a continuous set of states and the gradient of the cost function. In particular, we implement an optimal Universal-Not (Unot) gate \cite{Buzcircek2000, DeMartini2002, Ricci2004} with a variational circuit.

An Unot gate is defined as a gate that transforms a qubit in a particular state $\ket{\psi} = \cos(\theta/2)e^{i\phi} \ket{0}+\sin(\theta/2)\ket{1}$ into the orthogonal one $\ket{\psi^\perp} = \sin(\theta/2)e^{i\phi} \ket{0}-\cos(\theta/2) \ket{1}$. Since the transformation is not unitary, it can not be implemented deterministically; however it is possible to create an approximate Unot gate. In literature, a concept of \textit{optimal} Unot gate has been defined, as the unitary transformation which has the maximal average fidelity over all the possible input states. In Ref.~\cite{Buzcircek2000} they show that the maximal average fidelity achievable quantum mechanically is $F_\text{Unot} = 2/3$.

To provide a variational implementation of an optimal Unot gate, we use a universal three-mode interferometer inside the six-mode one, where the qubit is encoded with a dual rail encoding in the first two modes of the interferometer and the output state is post-selected on the condition that no photon is measured at the output of the third mode (see Fig.~\ref{fig:Universal-not}a).
To train the circuit we chose as input, a state defined by the two angles $\ket{\Psi(\theta_p, \phi_p)}$ and we directly measure the output state fidelity by projecting the final dual-rail encoded state on the expected orthogonal output state $\ket{\Psi^\bot (\theta_p, \phi_p)}$.
The cost function employed to train the circuit is defined as:
\begin{equation}
\begin{split}
    S(\Vec{\theta}) &= -\int_{0}^{\pi}\int_0^{2\pi} \frac{\sin\theta_p \dd \theta_p\dd \phi_p}{4\pi} F(\theta_p,\phi_p, \Vec{\theta}) P_s(\theta_p,\phi_p, \Vec{\theta}) \\
    & = -\int_{0}^{\pi}\int_0^{2\pi} \frac{\sin\theta_p \dd \theta_p\dd \phi_p}{4\pi} P_1(\theta_p,\phi_p, \Vec{\theta})
\end{split}
\end{equation}
where $F(\theta_p,\phi_p, \Vec{\theta})$ is the fidelity, $P_s(\theta_p,\phi_p, \Vec{\theta})$ is the success probability of the protocol, $P_1 (\theta_p,\phi_p, \Vec{\theta})$ is the probability that the photon came out from the first mode after the projection that is exactly the product of the previous two, and $\Vec{\theta}$ are the variational parameters of the Unot circuit.  The proposed cost function is designed to maximise the output fidelity, favouring the states with a higher success probability.
Such an average over the Bloch sphere can be estimated through the parameter shift rule for the integrals. In particular, we found  that the calculation of the cost function is equivalent to estimate $S(\Vec{\theta})$ as
\begin{equation}
    S(\Vec{\theta})=-\frac{1}{6}\sum_{(i,j) \in Q} P_1(i,j, \Vec{\theta})
\end{equation}
where $Q$ is a set of six pairs of angles $(\theta_p, \phi_p)$. More precisely $Q = \{(0,0), (\frac{\pi}{2}, 0), (\frac{\pi}{2}, \frac{\pi}{2}), (\frac{\pi}{2},\pi), (\frac{\pi}{2}, \frac{3}{2}\pi), (\pi, 0)\}$. The complete derivation is presented in the Appendix E.

Then, we move to the experimental minimization of the cost function that is performed via the 1-photon rule in the 6-mode photonic chip. The circuit is programmed as in Fig.~\ref{fig:Universal-not}a, where the phases $\Vec{\theta}=\{\theta_1, \dots \theta_5\}$ are the variational phases, the blue and the purple phases are the one used for the preparation and the projection on the orthogonal state. The minimization process follows the same procedure as the previous section. The quantum hardware computes the cost function gradient that is fed into the minimization routine based on the BFGS method. This time the quantum circuit also computes the cost function.

The results are summarised in the inset of Fig.~\ref{fig:Universal-not}b.
the final result of the minimization is $S_f = -0.654\pm 0.008$ which is compatible with a theoretical result of  $S_t = -0.666$.
To test the results we compute the average fidelity over a test set of $1000$ qubit states uniformly sampled on the Block sphere. We obtained a result of $\langle F \rangle = 0.660 \pm 0.005$ that is compatible with the results of an optimal Unot gate $F_{Unot} = 2/3 = 0.666$. Furthermore, the accordance with the expectation is confirmed by the fidelity distributions reported in Fig. \ref{fig:Universal-not}b.

\section{Discussion}
In this work, we presented and experimentally validated an extension of the parameter shift rule typical of the circuit-based model of quantum computation to a photonic-based formalism. 
To this end, we showed that the functional relationship between the probabilities at the output of a linear optical interferometer and the internal phases which define an implemented unitary transformation can be expressed as a trigonometric series whose degree depends on the number of photons at the input of the interferometer. These results enable the theory of trigonometric interpolation to be applied in order to reconstruct the function, its derivative and its integral by evaluating the function itself in a finite number of points uniformly distributed around a given point $x_0$.
Similar results for the computation of the gradient in a linear optical network are obtained in concurrent works \cite{facelli2024exactgradientslinearoptics, pappalardo2024photonicparametershiftruleenabling} employing an alternative formalism compared to the present work.
Moreover, with our formalism, we showed that the main sources of noise in a photonic experiment such as losses and partial distinguishability and mixedness of the state do not affect the functional form of the observables. This allows the possibility to apply the parameter shift rule without any particular change even in noisy conditions or without the necessity of a full characterization of the photonic apparatus losses and source indistinguishability. 
We also demonstrated that it is possible to reduce the number of measurements required for such computations by exploiting the regular geometry of a universal optical interferometer. Indeed, we proved that the degree of the trigonometric function is related to the number of photons propagating simultaneously through the phase under consideration.

The analytical theory developed here was then tested experimentally, exploiting a hybrid-photonic platform comprising of a spontaneous parametric down-conversion two-photon source and a 6-mode fully reconfigurable interferometer. In particular, we first validated the possibility of computing on-device both derivatives and integrals of cost functions, highlighting in the process how it is possible to reduce the number of measurements required using the information on the particular structure of the interferometer and on the relationship between input and output state.
Subsequently, the parameter shift rule was exploited to tackle two different variational optimisation problems. In particular, we showed how the gradients computed via the parameter shift rule can be fed to a gradient-based classical optimization algorithm to implement a variational eigensolver. This allows us to iteratively find the optimal circuit parameters corresponding to a chosen task. Here, we demonstrated its functioning to estimate the ground state energy of a hydrogen molecule. Then, we employed a similar approach to implement an optimal universal not gate on our photonic platform, achieving variationally the maximum average fidelity expected within quantum mechanics via a variational approach.

Overall, we show the possibility of computing the derivati

ve and the integral with respect to a parameter in an arbitrary linear optical interferometer.
In contrast to previous proposals \cite{defelice2024}, the algorithm requires neither more photons nor more modes than those needed to evaluate the cost function.
These results can be readily extended to more complex architectures, with a higher number of photons or optical modes, thus paving the way for further developments of efficient variational algorithms based on the use of reconfigurable integrated photonic circuits.
As a future perspective, one can try to further optimise the algorithm by exploiting the geometry of the interferometer under consideration and using classically inspired techniques such as back-propagation that have already found application in the circuit-based paradigm \cite{bowles2024,abbas2023}.

\section{Acknowledgements}
This work is supported by the European Union via project QLASS (“Quantum Glass-based Photonic Integrated Circuits” - Grant Agreement No. 101135876)  and from PNRR MUR project
PE0000023-NQSTI (Spoke 4).
The photonic chip was partially fabricated at PoliFAB, the micro- and nanofabrication facility of Politecnico di Milano (\href{https://www.polifab.polimi.it/}{https://www.polifab.polimi.it/}). 


\clearpage
\appendix

\section{Maximum number reduction}
In this section, we demonstrate the relation between the number of photons that pass through the phase of interest and the degree of the trigonometric function that represents the probability of a particular input-output configuration.

Recovering the demonstration in the main text we can say that 
\begin{equation}
    U_{hj}(\varphi) = \sum_{k} U^{(2)}_{hk} U^{(1)}_{kj} e^{i\delta_{kr}\varphi} \coloneqq c_{hj} + d_{hj} e^{i \varphi}
\end{equation}
However, by refining the reasoning, we can say that if there are no paths connecting the input $j$  and the output $h$ passing through the mode $r$ then the coefficient $d_{hj}$ is null.
This allows us to revise the equation { Eq.~13} of the main text 
\begin{equation}
    \prod_{h = 1}^N U_{\sigma(\Vec{d}(\Vec{s})_h), \Vec{d}(\Vec{r})_h}  = \sum_{k=0}^{K_{\sigma}} c_k^{(\sigma)} e^{ik\varphi}
\end{equation}
with an upper bound of the degree of the equation not anymore $N$ but a value $K_\sigma\leq N$ that counts the number of non-null coefficients $d_{hj}$, i.e. the maximum number of photons that can pass through the phase on the mode $r$ at the same time. Obviously, the upper bound $K_\sigma$ depends both on the particular permutation $\sigma$ considered and on the input and output states, omitted in the notation for readability reasons.
Defined $K_M = \max_{\sigma} K_\sigma$ the maximum degree for all the possible permutations, i.e. the maximum number of photons that can pass through the phase at the same time, then the upper bound for { Eq.~14} and {10} instead of $N$ became $K_M$ proving that
\begin{equation}
   P(\Vec{s}\,|\Vec{r}) = \sum_{k=0}^{K_M} a_k cos(k\varphi) + b_k sin(k \varphi)
\end{equation}

\section{Integral computation}
In this section, we demonstrate the formula for the average of the functions.
The goal is to compute the integral
\begin{equation}
    M = \int_{-\pi}^{\pi} \dd x f(x) g(x) 
\end{equation}
since the function $f(x)$ is a Fourier series of degree $R$ then we can use the Dirichlet kernel \cite{trigonometric2015} to rewrite the function as
\begin{equation}
    f(x) = \sum_{k = 0}^{2R} f\left(\frac{2k}{2R+1}\pi\right) \frac{(-1)^K}{2R+1}\frac{\sin(\frac{2R+1}{2}x)}{\sin(\frac{x}{2}-\frac{k}{2R+1}\pi)}
    \label{eq:kernel}
\end{equation}
this equation shape gives us the final results presented in the main text.

\subsection{odd function}
If the distribution function is odd then we can rewrite the original integral as
\begin{equation}
\begin{aligned}
    M &= \int_{-\pi}^{\pi} \dd x \frac{f(x) g(x) +f(-x)g(-x)}{2} \\
    &= \int_{-\pi}^{\pi} \dd x \frac{f(x)- f(-x)}{2} g(x)
\end{aligned}
\end{equation}
we can define $f_o(x)$ is the odd component of the function $f(x)$ as $f_o(x) = \frac{f(x)- f(-x)}{2}$.
As previously done we can rewrite the odd component of the function as \cite{Wierichs2022}
\begin{equation}
    f_o(x) = \sum_{k = 1}^R f_o\left(\frac{2k-1}{2R}\pi\right)\frac{(-1)^k}{R}\frac{cos(Rx)sin(x)}{cos(\frac{2k-1}{2R}\pi)-cos(x)}
\end{equation}
so defining the coefficients
\begin{equation}
    c_k = \int_{-\pi}^{\pi} \dd x \; g(x)\frac{cos(Rx)sin(x)}{cos(\frac{2k-1}{2R}\pi)-cos(x)}
\end{equation}
we can rewrite the original integral as
\begin{equation}
    M = \sum_{k = 1}^R \frac{c_k(-1)^k}{2R} \left[f\left(\frac{2k-1}{2R}\pi\right)-f\left(-\frac{2k-1}{2R}\pi\right)\right]
\end{equation}
which require $2R$ evaluation of the function $f(x)$ instead of $2R+1$ of the previous method.

\subsection{Even functions}
Similar to the previous case if the distribution function is even then the integral can be rewritten as 
\begin{equation}
\begin{aligned}
    M &= \int_{-\pi}^{\pi} \dd x \frac{f(x) g(x) +f(-x)g(-x)}{2} \\
    &= \int_{-\pi}^{\pi} \dd x \frac{f(x)+ f(-x)}{2} g(x)
    \end{aligned}
\end{equation}
defining $f_e(x)$ the even component of the function $f(x)$ as $f_e(x) = \frac{f(x)+ f(-x)}{2}$.
we can rewrite the function as\cite{Wierichs2022}
\begin{equation}
    f_e(x) = \sum_{k = 0}^{2R-1} f\left(\frac{k\pi}{R}\right) \frac{(-1)^k}{2R}\frac{sin(Rx)sin(x)}{cos(\frac{k\pi}{R})-cos(x)}
\end{equation}
defining
\begin{equation}
    c_k = \int_{-\pi}^{\pi} \dd x g(x) \frac{sin(Rx) sin(x)}{cos(\frac{k\pi}{R})-cos(x)}
\end{equation}
this gives us the final expression for the original integral as
\begin{equation}
    M = \sum_{k = 0}^{2R-1} \frac{(-1)^k c_k}{2R} f\left(\frac{k\pi}{R}\right)
\end{equation}
with require $2R$ evaluation of the function compared to the $2R+1$ of the initial method

\section{Resource comparison}
In this Section, we provide a quantitative insight into the resources required by our algorithms compared to other algorithms employed in the literature.

First, we compare our algorithm with others in the literature. The papers \cite{facelli2024exactgradientslinearoptics} and \cite{pappalardo2024photonicparametershiftruleenabling} derive the same technique discussed in this manuscript with a different approach and formalism. Overall, this means that our work and \cite{facelli2024exactgradientslinearoptics, pappalardo2024photonicparametershiftruleenabling} are to be considered equivalent from both performance and resource expenditure point of view.
In contrast, in Ref. \cite{defelice2024}, they use a method that requires an interferometer of double the size compared to the one used to evaluate the cost function, and an additional photon. Moreover, the number of measurements scales
quadratically with the number of modes, while in our approach they scale linearly with the number of photons at the input of the interferometer. In most cases, the number of photons is lower than the number of modes, meaning that our approach is more efficient in terms of the required measure.

Second, we want to compare the application of our algorithm in an optimization scenario. The comparison of gradient-based and gradient-free algorithms in terms of the required number of function evaluations is an open problem since it depends on the specific problem, in particular on the number of parameters to be optimized, on the properties of the function and on eventual bounds on the parameter space.

For example, let us consider the variational implementation of a Universal-not gate, which we present in Section III C. We can analyze the required number of function evaluations for the training of the variational circuit as a function of the size of the interferometer, i.e. the number of variational parameters in the interferometer. In particular, we compare the BFGS algorithm (the gradient-based algorithm used in the experiment) and the Nelder-Mead algorithm, which is the one typically chosen among the gradient-free methods. The results of this comparison are summarized in Table~\ref{tab:Min_perform} where the average function evaluation is estimated over $10000$ trials with a random starting point in the hypercube $[0,2\pi)^{n}$. As we can see, except for the case with $n= 5$ parameters, the use of the gradient-based algorithm is preferable.
This shows that as the number of parameters increases, the gradient-based algorithm seems preferable to the gradient-free case, demonstrating the potential of the algorithm proposed in the main text.

\begin{table}[ht!]
    \centering
    \begin{tabular}{cccc}
    \toprule
        Modes & Parameters &\multicolumn{2}{c}{Averege functions evaluations}\\
        $(m)$ & $(n)$ & BFGS & Nelder-Mead \\
     \midrule
        $3$ & $5$ & $281\pm 66$ & $254 \pm 77$\\
        $3$ & $9$ & $299\pm 97$ & $347 \pm 120$\\
        $4$ & $16$ & $460\pm 150$ & $560 \pm 224$ \\
        $5$ & $25$ & $546 \pm 181$ & $674\pm 254$\\
        $6$ & $36$ & $718 \pm 233$ & $867\pm 231$\\
    \bottomrule
    \end{tabular}
    \caption{\textbf{Average function evaluations:} We report the average function evaluations for the training of a Variational Universal-not gate using two different algorithms, namely BFGS and Nelder-Mead. The mean and the standard deviation is estimated over $10000$ rials with a random starting point in the hypercube $[0,2\pi)^{n}$. for different number of modes $m$ and variational parameters $n$. Except for the first case where we use the interferometer depicted in Fig.~4a of the main text, we use the complete layout of Ref. \cite{Clements2016} that has $n = m^2$.}
    \label{tab:Min_perform}
\end{table}

\section{Hamiltonian minimization protocol}

For each measurement, two indistinguishable photons enter in modes 3 and 5 of the chip and undergo a four-mode unitary transformation which gives the state $\ket{\Phi}$. Then $\bra{\Psi} H\ket{\Psi}$ is measured by measuring the probabilities of each term of Eq.~{ 19} of the maintext. Considering we're in the Z base for the two qubits, we can measure:
\begin{center}
    \begin{tabular}{c|c}
    $P(II)$  &  $P(ZI)$\\
    $P_{13} + P_{14} + P_{23} + P_{24}~~$ & $~~P_{13} - P_{14} + P_{23} - P_{24}~~$\\
    \hline
    $P(IZ)$ & $P(ZZ)$\\
    $~~P_{13} + P_{14} - P_{23} - P_{24}~~$ & $~~P_{13} - P_{14} - P_{23} + P_{24}$
    \end{tabular}
\end{center}
where 
\begin{equation}
    P_{ij} = \frac{N_{ij}}{N_{13}+N_{14}+N_{23}+N_{24}}
\end{equation}

By applying a base change matrix from Z to X before the measurement, which in the chip is a $\pi/2$ beamsplitter (red area in Fig.~{ 3}  of the maintext), we can also measure:
\begin{center}
    $P(XX) = P_{13} - P_{14} - P_{23} + P_{24}$
\end{center}

Note that the probability here is not the global probability of success, but the probability postselected on the event that un photon is in the first two modes and the sencod photon in the last two modes, i.e. that we generate two proper qubits. 
Since the coincidences $N_{ij}$ follow the 2 photons rule, as they are proportional to the output probabilities, we can use the chain rule to compute the partial derivatives of the probabilities $P_{ij}$ as 

\begin{equation}
\begin{aligned}
    \frac{\partial P_{ij}}{\partial \theta_k} &= \frac{\partial}{\partial \theta_k}\frac{N_{ij}}{N_{13}+N_{14}+N_{23}+N_{24}} \\
    &= \frac{1}{S}\frac{\partial N_{ij}}{\partial\theta_k} + \frac{N_{ij}}{(S)^2}\frac{\partial S}{\partial \theta_k}
\end{aligned}
\end{equation}

where we define $S=N_{13}+N_{14}+N_{23}+N_{24}$ and every partial derivative of the counts computed thanks to the two photons rule

\begin{equation}
\begin{split}
    \frac{\partial N_{ij}}{\partial\theta_k} = &\frac{2+\sqrt{2}}{4}\left[N_{ij}\left(\theta_k+\frac{\pi}{4}\right)-N_{ij}\left(\theta_k-\frac{\pi}{4}\right)\right]+\\
    -&\frac{2-\sqrt{2}}{4}\left[N_{ij}\left(\theta_k+\frac{3\pi}{4}\right)-N_{ij}\left(\theta_k-\frac{3\pi}{4}\right)\right]
\end{split}
\end{equation}

\section{Unot cost function}
In this section, we show how to apply the parameter shift rule to compute the cost function of the Unot problem.
We define $P_1(\theta_p, \phi_p, \theta_m, \phi_m, \Vec{\theta}\;)$ as the probability that the photon came out from the first mode of the interferometer as a function to the phases in the preparation stage and in the measurement one respectively.
Due to the particular form of the interferometer, the function presents some useful symmetries
\begin{equation}
\begin{aligned}
    P_1(2\pi-\theta_p, \phi_p, \theta_m, \phi_m, \Vec{\theta}\;) &= P_1(\theta_p, \phi_p+\pi, \theta_m, \phi_m, \Vec{\theta}\;)\\
    P_1(\theta_p, \phi_p, 2\pi-\theta_m, \phi_m, \Vec{\theta}\;) &= P_1(\theta_p, \phi_p+\pi, \theta_m, \phi_m, \Vec{\theta}\;)\\
    P_1(0, \phi_p, \theta_m, \phi_m, \Vec{\theta}\;) &= P_1(0, 0, \theta_m, \phi_m, \Vec{\theta}\;)\\
    P_1(\theta_p, \phi_p, 0, \phi_m, \Vec{\theta}\;) &= P_1(\theta_p, \phi_p, 0,0, \Vec{\theta}\;)\\
    P_1(\pi, \phi_p, \theta_m, \phi_m, \Vec{\theta}\;) &= P_1(\pi, 0, \theta_m, \phi_m, \Vec{\theta}\:)\\
    P_1(\theta_p, \phi_p, \pi, \phi_m, \Vec{\theta}\;) &= P_1(\theta_p, \phi_p, \pi,0, \Vec{\theta}\;)
\end{aligned}
\end{equation}

Starting from the cost function
\begin{equation}
    S(\Vec{\theta}\;) = -\int_{0}^{\pi}\int_0^{2\pi} \frac{sin\theta_p \dd \theta_p\dd\phi_p}{4\pi} P_1(\theta,\phi,\pi-\theta, \pi-\phi, \Vec{\theta}\;)
\end{equation}
Since only one photon are injected in the circuit then the probability $P_1$ is a trigonometric function of degree $1$ in all four phases.
However, since we control two phases simultaneously this implies that in the variable $\theta_p$ and $\phi_p$ the probability function is a trigonometric function of degree two that for more clarity we write as $P_1(\theta,\phi, \Vec{\theta}\;) = P_1(\theta,\phi,\pi-\theta, \pi-\phi, \Vec{\theta}\;)$.
Now that we fin the value of $R$, we need to extend the integration over $\theta$ from $[0,\pi]$ to $[0,2\pi]$. Using the symmetries and the periodicity of the function we can write
\begin{equation}
\begin{split}
    &\int_{0}^{\pi}\int_0^{2\pi} \frac{sin\theta_p \dd \theta_p}{2} \frac{\dd \phi_p}{2\pi} P_1(\theta_,\phi_p, \Vec{\theta}\;) = \\
    &= \int_{0}^{\pi}\int_0^{2\pi} \frac{sin\theta_p \dd \theta_p}{2} \frac{\dd \phi_p}{2\pi} P_1(\theta_p,\phi_p+\pi, \Vec{\theta}\;) \\
    &= \int_{0}^{\pi}\int_0^{2\pi} \frac{sin\theta_p \dd \theta_p}{2} \frac{\dd \phi_p}{2\pi} P_1(2\pi-\theta_p,\phi_p, \Vec{\theta}\;) \\
    &= \int_{2\pi}^{\pi}\int_0^{2\pi} \frac{-sin (2\pi-\theta_p) \dd \theta_p}{2} \frac{\dd \phi_p}{2\pi} P_1(\theta_p,\phi_p, \Vec{\theta}\;)\\
    &= -\int_{\pi}^{2\pi}\int_0^{2\pi} \frac{sin\theta_p \dd \theta_p}{2} \frac{\dd \phi_p}{2\pi} P_1(\theta_p,\phi_p, \Vec{\theta}\;)
\end{split}
\end{equation}
this implies that we can rewrite the cost function as
\begin{equation}
    S(\Vec{\theta}\;) = -\int_{0}^{2\pi}\int_0^{2\pi} \frac{\abs{sin\theta_p} \dd \theta_p}{4} \frac{\dd \phi_p}{2\pi} P_1(\theta_p,\phi_p, \Vec{\theta}\;)
\end{equation}

Since both $g(\theta_p) = \frac{\abs{sin\theta_p}}{4}$ and $h(\phi) = \frac{1}{2\pi}$ are even functions we can calculate the coefficient for the corresponding parameter shift rules.
\begin{align}
    & \begin{split}
    c_k = \int_{0}^{2\pi} \dd \theta_p \frac{\abs{sin \theta_p}}{4} \frac{sin(2\theta_p) sin(\theta_p)}{cos(\frac{k\pi}{2})-cos(\theta_p)} \\
    c_0 = {\phantom -}\frac{2}{3} \quad 
    c_1 = -\frac{4}{3} \quad 
    c_2 = {\phantom -}\frac{2}{3}\ \quad 
    c_3 = -\frac{4}{3}
    \end{split} \\
    \\
    & \begin{split}
        d_k = \int_{0}^{2\pi} \dd \phi_p \frac{1}{2\pi} \frac{sin(2\phi_p) sin(\phi_p)}{cos(\frac{k\pi}{2})-cos(\phi_p)} \\
        d_0 = {\phantom -}1 \quad 
        d_1 = -1 \quad 
        d_2 = {\phantom -}1 \quad 
        d_3 = -1
    \end{split}
\end{align}

Giving the expression
\begin{equation}
    S(\Vec{\theta}\;) = -\sum_{h = 0, k = 0}^{3,3} \frac{(-1)^{h+k}c_h,d_k}{16} P_1\left(\frac{h\pi}{2},\frac{k\pi}{2}, \Vec{\theta}\right)
\end{equation}
This expression requires $16$ function evaluations. However, it is possible to use the symmetries of the probability function to reduce the number of evaluations.
In particular
\begin{equation}
\begin{split}
S(\Vec{\theta}\;) = - \biggl[\frac{1}{24} \sum_{k = 0}^3 P_1\left(0,\frac{k\pi}{2}, \Vec{\theta}\right)+\frac{1}{12} \sum_{k = 0}^3 P_1\left(\frac{\pi}{2},\frac{k\pi}{2}, \Vec{\theta}\right)+\\
+\frac{1}{24} \sum_{k = 0}^3 P_1\left(\pi,\frac{k\pi}{2}, \Vec{\theta}\right)+\frac{1}{12} \sum_{k = 0}^3 P_1\left(\frac{3\pi}{2},\frac{k\pi}{2}, \Vec{\theta}\right)\biggr]\\
\; = - \left[\frac{1}{6} P_1\left(0,0, \Vec{\theta}\right)+\frac{1}{6} \sum_{k = 0}^3 P_1\left(\frac{\pi}{2},\frac{k\pi}{2}, \Vec{\theta}\right)+\frac{1}{6}  P_1\left(\pi,0, \Vec{\theta}\right)\right]
\end{split}
\end{equation}
giving the final result present in the main text.

\clearpage

\begin{thebibliography}{47}%
	\makeatletter
	\providecommand \@ifxundefined [1]{%
		\@ifx{#1\undefined}
	}%
	\providecommand \@ifnum [1]{%
		\ifnum #1\expandafter \@firstoftwo
		\else \expandafter \@secondoftwo
		\fi
	}%
	\providecommand \@ifx [1]{%
		\ifx #1\expandafter \@firstoftwo
		\else \expandafter \@secondoftwo
		\fi
	}%
	\providecommand \natexlab [1]{#1}%
	\providecommand \enquote  [1]{``#1''}%
	\providecommand \bibnamefont  [1]{#1}%
	\providecommand \bibfnamefont [1]{#1}%
	\providecommand \citenamefont [1]{#1}%
	\providecommand \href@noop [0]{\@secondoftwo}%
	\providecommand \href [0]{\begingroup \@sanitize@url \@href}%
	\providecommand \@href[1]{\@@startlink{#1}\@@href}%
	\providecommand \@@href[1]{\endgroup#1\@@endlink}%
	\providecommand \@sanitize@url [0]{\catcode `\\12\catcode `\$12\catcode `\&12\catcode `\#12\catcode `\^12\catcode `\_12\catcode `\%12\relax}%
	\providecommand \@@startlink[1]{}%
	\providecommand \@@endlink[0]{}%
	\providecommand \url  [0]{\begingroup\@sanitize@url \@url }%
	\providecommand \@url [1]{\endgroup\@href {#1}{\urlprefix }}%
	\providecommand \urlprefix  [0]{URL }%
	\providecommand \Eprint [0]{\href }%
	\providecommand \doibase [0]{https://doi.org/}%
	\providecommand \selectlanguage [0]{\@gobble}%
	\providecommand \bibinfo  [0]{\@secondoftwo}%
	\providecommand \bibfield  [0]{\@secondoftwo}%
	\providecommand \translation [1]{[#1]}%
	\providecommand \BibitemOpen [0]{}%
	\providecommand \bibitemStop [0]{}%
	\providecommand \bibitemNoStop [0]{.\EOS\space}%
	\providecommand \EOS [0]{\spacefactor3000\relax}%
	\providecommand \BibitemShut  [1]{\csname bibitem#1\endcsname}%
	\let\auto@bib@innerbib\@empty
	\bibitem [{\citenamefont {Cerezo}\ \emph {et~al.}(2021)\citenamefont {Cerezo}, \citenamefont {Arrasmith}, \citenamefont {Babbush}, \citenamefont {Benjamin}, \citenamefont {Endo}, \citenamefont {Fujii}, \citenamefont {McClean}, \citenamefont {Mitarai}, \citenamefont {Yuan}, \citenamefont {Cincio},\ and\ \citenamefont {Coles}}]{Cerezo2021}%
	\BibitemOpen
	\bibfield  {author} {\bibinfo {author} {\bibfnamefont {M.}~\bibnamefont {Cerezo}}, \bibinfo {author} {\bibfnamefont {A.}~\bibnamefont {Arrasmith}}, \bibinfo {author} {\bibfnamefont {R.}~\bibnamefont {Babbush}}, \bibinfo {author} {\bibfnamefont {S.~C.}\ \bibnamefont {Benjamin}}, \bibinfo {author} {\bibfnamefont {S.}~\bibnamefont {Endo}}, \bibinfo {author} {\bibfnamefont {K.}~\bibnamefont {Fujii}}, \bibinfo {author} {\bibfnamefont {J.~R.}\ \bibnamefont {McClean}}, \bibinfo {author} {\bibfnamefont {K.}~\bibnamefont {Mitarai}}, \bibinfo {author} {\bibfnamefont {X.}~\bibnamefont {Yuan}}, \bibinfo {author} {\bibfnamefont {L.}~\bibnamefont {Cincio}},\ and\ \bibinfo {author} {\bibfnamefont {P.~J.}\ \bibnamefont {Coles}},\ }\bibfield  {title} {\bibinfo {title} {Variational quantum algorithms},\ }\href {https://doi.org/10.1038/s42254-021-00348-9} {\bibfield  {journal} {\bibinfo  {journal} {Nature Reviews Physics}\ }\textbf {\bibinfo {volume} {3}},\ \bibinfo {pages} {625–644} (\bibinfo {year} {2021})}\BibitemShut
	{NoStop}%
	\bibitem [{\citenamefont {Bharti}\ \emph {et~al.}(2022)\citenamefont {Bharti}, \citenamefont {Cervera-Lierta}, \citenamefont {Kyaw}, \citenamefont {Haug}, \citenamefont {Alperin-Lea}, \citenamefont {Anand}, \citenamefont {Degroote}, \citenamefont {Heimonen}, \citenamefont {Kottmann}, \citenamefont {Menke}, \citenamefont {Mok}, \citenamefont {Sim}, \citenamefont {Kwek},\ and\ \citenamefont {Aspuru-Guzik}}]{Rev_Noisy_Var}%
	\BibitemOpen
	\bibfield  {author} {\bibinfo {author} {\bibfnamefont {K.}~\bibnamefont {Bharti}}, \bibinfo {author} {\bibfnamefont {A.}~\bibnamefont {Cervera-Lierta}}, \bibinfo {author} {\bibfnamefont {T.~H.}\ \bibnamefont {Kyaw}}, \bibinfo {author} {\bibfnamefont {T.}~\bibnamefont {Haug}}, \bibinfo {author} {\bibfnamefont {S.}~\bibnamefont {Alperin-Lea}}, \bibinfo {author} {\bibfnamefont {A.}~\bibnamefont {Anand}}, \bibinfo {author} {\bibfnamefont {M.}~\bibnamefont {Degroote}}, \bibinfo {author} {\bibfnamefont {H.}~\bibnamefont {Heimonen}}, \bibinfo {author} {\bibfnamefont {J.~S.}\ \bibnamefont {Kottmann}}, \bibinfo {author} {\bibfnamefont {T.}~\bibnamefont {Menke}}, \bibinfo {author} {\bibfnamefont {W.-K.}\ \bibnamefont {Mok}}, \bibinfo {author} {\bibfnamefont {S.}~\bibnamefont {Sim}}, \bibinfo {author} {\bibfnamefont {L.-C.}\ \bibnamefont {Kwek}},\ and\ \bibinfo {author} {\bibfnamefont {A.}~\bibnamefont {Aspuru-Guzik}},\ }\bibfield  {title} {\bibinfo {title} {Noisy intermediate-scale quantum algorithms},\ }\href
	{https://doi.org/10.1103/RevModPhys.94.015004} {\bibfield  {journal} {\bibinfo  {journal} {Rev. Mod. Phys.}\ }\textbf {\bibinfo {volume} {94}},\ \bibinfo {pages} {015004} (\bibinfo {year} {2022})}\BibitemShut {NoStop}%
	\bibitem [{\citenamefont {Endo}\ \emph {et~al.}(2020)\citenamefont {Endo}, \citenamefont {Sun}, \citenamefont {Li}, \citenamefont {Benjamin},\ and\ \citenamefont {Yuan}}]{Endo2020}%
	\BibitemOpen
	\bibfield  {author} {\bibinfo {author} {\bibfnamefont {S.}~\bibnamefont {Endo}}, \bibinfo {author} {\bibfnamefont {J.}~\bibnamefont {Sun}}, \bibinfo {author} {\bibfnamefont {Y.}~\bibnamefont {Li}}, \bibinfo {author} {\bibfnamefont {S.~C.}\ \bibnamefont {Benjamin}},\ and\ \bibinfo {author} {\bibfnamefont {X.}~\bibnamefont {Yuan}},\ }\bibfield  {title} {\bibinfo {title} {Variational quantum simulation of general processes},\ }\href {https://doi.org/10.1103/physrevlett.125.010501} {\bibfield  {journal} {\bibinfo  {journal} {Physical Review Letters}\ }\textbf {\bibinfo {volume} {125}},\ \bibinfo {pages} {010501} (\bibinfo {year} {2020})}\BibitemShut {NoStop}%
	\bibitem [{\citenamefont {Yuan}\ \emph {et~al.}(2019)\citenamefont {Yuan}, \citenamefont {Endo}, \citenamefont {Zhao}, \citenamefont {Li},\ and\ \citenamefont {Benjamin}}]{Yuan2019}%
	\BibitemOpen
	\bibfield  {author} {\bibinfo {author} {\bibfnamefont {X.}~\bibnamefont {Yuan}}, \bibinfo {author} {\bibfnamefont {S.}~\bibnamefont {Endo}}, \bibinfo {author} {\bibfnamefont {Q.}~\bibnamefont {Zhao}}, \bibinfo {author} {\bibfnamefont {Y.}~\bibnamefont {Li}},\ and\ \bibinfo {author} {\bibfnamefont {S.~C.}\ \bibnamefont {Benjamin}},\ }\bibfield  {title} {\bibinfo {title} {Theory of variational quantum simulation},\ }\href {https://doi.org/10.22331/q-2019-10-07-191} {\bibfield  {journal} {\bibinfo  {journal} {Quantum}\ }\textbf {\bibinfo {volume} {3}},\ \bibinfo {pages} {191} (\bibinfo {year} {2019})}\BibitemShut {NoStop}%
	\bibitem [{\citenamefont {Cao}\ \emph {et~al.}(2019)\citenamefont {Cao}, \citenamefont {Romero}, \citenamefont {Olson}, \citenamefont {Degroote}, \citenamefont {Johnson}, \citenamefont {Kieferová}, \citenamefont {Kivlichan}, \citenamefont {Menke}, \citenamefont {Peropadre}, \citenamefont {Sawaya}, \citenamefont {Sim}, \citenamefont {Veis},\ and\ \citenamefont {Aspuru-Guzik}}]{Cao2019}%
	\BibitemOpen
	\bibfield  {author} {\bibinfo {author} {\bibfnamefont {Y.}~\bibnamefont {Cao}}, \bibinfo {author} {\bibfnamefont {J.}~\bibnamefont {Romero}}, \bibinfo {author} {\bibfnamefont {J.~P.}\ \bibnamefont {Olson}}, \bibinfo {author} {\bibfnamefont {M.}~\bibnamefont {Degroote}}, \bibinfo {author} {\bibfnamefont {P.~D.}\ \bibnamefont {Johnson}}, \bibinfo {author} {\bibfnamefont {M.}~\bibnamefont {Kieferová}}, \bibinfo {author} {\bibfnamefont {I.~D.}\ \bibnamefont {Kivlichan}}, \bibinfo {author} {\bibfnamefont {T.}~\bibnamefont {Menke}}, \bibinfo {author} {\bibfnamefont {B.}~\bibnamefont {Peropadre}}, \bibinfo {author} {\bibfnamefont {N.~P.~D.}\ \bibnamefont {Sawaya}}, \bibinfo {author} {\bibfnamefont {S.}~\bibnamefont {Sim}}, \bibinfo {author} {\bibfnamefont {L.}~\bibnamefont {Veis}},\ and\ \bibinfo {author} {\bibfnamefont {A.}~\bibnamefont {Aspuru-Guzik}},\ }\bibfield  {title} {\bibinfo {title} {Quantum chemistry in the age of quantum computing},\ }\href {https://doi.org/10.1021/acs.chemrev.8b00803} {\bibfield
		{journal} {\bibinfo  {journal} {Chemical Reviews}\ }\textbf {\bibinfo {volume} {119}},\ \bibinfo {pages} {10856–10915} (\bibinfo {year} {2019})}\BibitemShut {NoStop}%
	\bibitem [{\citenamefont {Grimsley}\ \emph {et~al.}(2019)\citenamefont {Grimsley}, \citenamefont {Economou}, \citenamefont {Barnes},\ and\ \citenamefont {Mayhall}}]{Grimsley2019}%
	\BibitemOpen
	\bibfield  {author} {\bibinfo {author} {\bibfnamefont {H.~R.}\ \bibnamefont {Grimsley}}, \bibinfo {author} {\bibfnamefont {S.~E.}\ \bibnamefont {Economou}}, \bibinfo {author} {\bibfnamefont {E.}~\bibnamefont {Barnes}},\ and\ \bibinfo {author} {\bibfnamefont {N.~J.}\ \bibnamefont {Mayhall}},\ }\bibfield  {title} {\bibinfo {title} {An adaptive variational algorithm for exact molecular simulations on a quantum computer},\ }\href {https://doi.org/10.1038/s41467-019-10988-2} {\bibfield  {journal} {\bibinfo  {journal} {Nature Communications}\ }\textbf {\bibinfo {volume} {10}},\ \bibinfo {pages} {3007} (\bibinfo {year} {2019})}\BibitemShut {NoStop}%
	\bibitem [{\citenamefont {Peruzzo}\ \emph {et~al.}(2014)\citenamefont {Peruzzo}, \citenamefont {McClean}, \citenamefont {Shadbolt}, \citenamefont {Yung}, \citenamefont {Zhou}, \citenamefont {Love}, \citenamefont {Aspuru-Guzik},\ and\ \citenamefont {O’Brien}}]{Peruzzo2014}%
	\BibitemOpen
	\bibfield  {author} {\bibinfo {author} {\bibfnamefont {A.}~\bibnamefont {Peruzzo}}, \bibinfo {author} {\bibfnamefont {J.}~\bibnamefont {McClean}}, \bibinfo {author} {\bibfnamefont {P.}~\bibnamefont {Shadbolt}}, \bibinfo {author} {\bibfnamefont {M.-H.}\ \bibnamefont {Yung}}, \bibinfo {author} {\bibfnamefont {X.-Q.}\ \bibnamefont {Zhou}}, \bibinfo {author} {\bibfnamefont {P.~J.}\ \bibnamefont {Love}}, \bibinfo {author} {\bibfnamefont {A.}~\bibnamefont {Aspuru-Guzik}},\ and\ \bibinfo {author} {\bibfnamefont {J.~L.}\ \bibnamefont {O’Brien}},\ }\bibfield  {title} {\bibinfo {title} {A variational eigenvalue solver on a photonic quantum processor},\ }\bibfield  {journal} {\bibinfo  {journal} {Nature Communications}\ }\textbf {\bibinfo {volume} {5}},\ \href {https://doi.org/10.1038/ncomms5213} {10.1038/ncomms5213} (\bibinfo {year} {2014})\BibitemShut {NoStop}%
	\bibitem [{\citenamefont {Santagati}\ \emph {et~al.}(2018)\citenamefont {Santagati}, \citenamefont {Wang}, \citenamefont {Gentile}, \citenamefont {Paesani}, \citenamefont {Wiebe}, \citenamefont {McClean}, \citenamefont {Morley-Short}, \citenamefont {Shadbolt}, \citenamefont {Bonneau}, \citenamefont {Silverstone}, \citenamefont {Tew}, \citenamefont {Zhou}, \citenamefont {O’Brien},\ and\ \citenamefont {Thompson}}]{Santagati2018}%
	\BibitemOpen
	\bibfield  {author} {\bibinfo {author} {\bibfnamefont {R.}~\bibnamefont {Santagati}}, \bibinfo {author} {\bibfnamefont {J.}~\bibnamefont {Wang}}, \bibinfo {author} {\bibfnamefont {A.~A.}\ \bibnamefont {Gentile}}, \bibinfo {author} {\bibfnamefont {S.}~\bibnamefont {Paesani}}, \bibinfo {author} {\bibfnamefont {N.}~\bibnamefont {Wiebe}}, \bibinfo {author} {\bibfnamefont {J.~R.}\ \bibnamefont {McClean}}, \bibinfo {author} {\bibfnamefont {S.}~\bibnamefont {Morley-Short}}, \bibinfo {author} {\bibfnamefont {P.~J.}\ \bibnamefont {Shadbolt}}, \bibinfo {author} {\bibfnamefont {D.}~\bibnamefont {Bonneau}}, \bibinfo {author} {\bibfnamefont {J.~W.}\ \bibnamefont {Silverstone}}, \bibinfo {author} {\bibfnamefont {D.~P.}\ \bibnamefont {Tew}}, \bibinfo {author} {\bibfnamefont {X.}~\bibnamefont {Zhou}}, \bibinfo {author} {\bibfnamefont {J.~L.}\ \bibnamefont {O’Brien}},\ and\ \bibinfo {author} {\bibfnamefont {M.~G.}\ \bibnamefont {Thompson}},\ }\bibfield  {title} {\bibinfo {title} {Witnessing eigenstates for quantum
			simulation of hamiltonian spectra},\ }\bibfield  {journal} {\bibinfo  {journal} {Science Advances}\ }\textbf {\bibinfo {volume} {4}},\ \href {https://doi.org/10.1126/sciadv.aap9646} {10.1126/sciadv.aap9646} (\bibinfo {year} {2018})\BibitemShut {NoStop}%
	\bibitem [{\citenamefont {Maring}\ \emph {et~al.}(2024)\citenamefont {Maring}, \citenamefont {Fyrillas}, \citenamefont {Pont}, \citenamefont {Ivanov}, \citenamefont {Stepanov}, \citenamefont {Margaria}, \citenamefont {Hease}, \citenamefont {Pishchagin}, \citenamefont {Lemaître}, \citenamefont {Sagnes}, \citenamefont {Au}, \citenamefont {Boissier}, \citenamefont {Bertasi}, \citenamefont {Baert}, \citenamefont {Valdivia}, \citenamefont {Billard}, \citenamefont {Acar}, \citenamefont {Brieussel}, \citenamefont {Mezher}, \citenamefont {Wein}, \citenamefont {Salavrakos}, \citenamefont {Sinnott}, \citenamefont {Fioretto}, \citenamefont {Emeriau}, \citenamefont {Belabas}, \citenamefont {Mansfield}, \citenamefont {Senellart}, \citenamefont {Senellart},\ and\ \citenamefont {Somaschi}}]{Maring2024}%
	\BibitemOpen
	\bibfield  {author} {\bibinfo {author} {\bibfnamefont {N.}~\bibnamefont {Maring}}, \bibinfo {author} {\bibfnamefont {A.}~\bibnamefont {Fyrillas}}, \bibinfo {author} {\bibfnamefont {M.}~\bibnamefont {Pont}}, \bibinfo {author} {\bibfnamefont {E.}~\bibnamefont {Ivanov}}, \bibinfo {author} {\bibfnamefont {P.}~\bibnamefont {Stepanov}}, \bibinfo {author} {\bibfnamefont {N.}~\bibnamefont {Margaria}}, \bibinfo {author} {\bibfnamefont {W.}~\bibnamefont {Hease}}, \bibinfo {author} {\bibfnamefont {A.}~\bibnamefont {Pishchagin}}, \bibinfo {author} {\bibfnamefont {A.}~\bibnamefont {Lemaître}}, \bibinfo {author} {\bibfnamefont {I.}~\bibnamefont {Sagnes}}, \bibinfo {author} {\bibfnamefont {T.~H.}\ \bibnamefont {Au}}, \bibinfo {author} {\bibfnamefont {S.}~\bibnamefont {Boissier}}, \bibinfo {author} {\bibfnamefont {E.}~\bibnamefont {Bertasi}}, \bibinfo {author} {\bibfnamefont {A.}~\bibnamefont {Baert}}, \bibinfo {author} {\bibfnamefont {M.}~\bibnamefont {Valdivia}}, \bibinfo {author} {\bibfnamefont {M.}~\bibnamefont
			{Billard}}, \bibinfo {author} {\bibfnamefont {O.}~\bibnamefont {Acar}}, \bibinfo {author} {\bibfnamefont {A.}~\bibnamefont {Brieussel}}, \bibinfo {author} {\bibfnamefont {R.}~\bibnamefont {Mezher}}, \bibinfo {author} {\bibfnamefont {S.~C.}\ \bibnamefont {Wein}}, \bibinfo {author} {\bibfnamefont {A.}~\bibnamefont {Salavrakos}}, \bibinfo {author} {\bibfnamefont {P.}~\bibnamefont {Sinnott}}, \bibinfo {author} {\bibfnamefont {D.~A.}\ \bibnamefont {Fioretto}}, \bibinfo {author} {\bibfnamefont {P.-E.}\ \bibnamefont {Emeriau}}, \bibinfo {author} {\bibfnamefont {N.}~\bibnamefont {Belabas}}, \bibinfo {author} {\bibfnamefont {S.}~\bibnamefont {Mansfield}}, \bibinfo {author} {\bibfnamefont {P.}~\bibnamefont {Senellart}}, \bibinfo {author} {\bibfnamefont {J.}~\bibnamefont {Senellart}},\ and\ \bibinfo {author} {\bibfnamefont {N.}~\bibnamefont {Somaschi}},\ }\bibfield  {title} {\bibinfo {title} {A versatile single-photon-based quantum computing platform},\ }\href {https://doi.org/10.1038/s41566-024-01403-4} {\bibfield
		{journal} {\bibinfo  {journal} {Nature Photonics}\ }\textbf {\bibinfo {volume} {18}},\ \bibinfo {pages} {603–609} (\bibinfo {year} {2024})}\BibitemShut {NoStop}%
	\bibitem [{\citenamefont {Hoch}\ \emph {et~al.}(2024)\citenamefont {Hoch}, \citenamefont {Rodari}, \citenamefont {Caruccio}, \citenamefont {Polacchi}, \citenamefont {Carvacho}, \citenamefont {Giordani}, \citenamefont {Doosti}, \citenamefont {Nicolau}, \citenamefont {Pentangelo}, \citenamefont {Piacentini}, \citenamefont {Crespi}, \citenamefont {Ceccarelli}, \citenamefont {Osellame}, \citenamefont {Galvão}, \citenamefont {Spagnolo},\ and\ \citenamefont {Sciarrino}}]{hoch2024variational}%
	\BibitemOpen
	\bibfield  {author} {\bibinfo {author} {\bibfnamefont {F.}~\bibnamefont {Hoch}}, \bibinfo {author} {\bibfnamefont {G.}~\bibnamefont {Rodari}}, \bibinfo {author} {\bibfnamefont {E.}~\bibnamefont {Caruccio}}, \bibinfo {author} {\bibfnamefont {B.}~\bibnamefont {Polacchi}}, \bibinfo {author} {\bibfnamefont {G.}~\bibnamefont {Carvacho}}, \bibinfo {author} {\bibfnamefont {T.}~\bibnamefont {Giordani}}, \bibinfo {author} {\bibfnamefont {M.}~\bibnamefont {Doosti}}, \bibinfo {author} {\bibfnamefont {S.}~\bibnamefont {Nicolau}}, \bibinfo {author} {\bibfnamefont {C.}~\bibnamefont {Pentangelo}}, \bibinfo {author} {\bibfnamefont {S.}~\bibnamefont {Piacentini}}, \bibinfo {author} {\bibfnamefont {A.}~\bibnamefont {Crespi}}, \bibinfo {author} {\bibfnamefont {F.}~\bibnamefont {Ceccarelli}}, \bibinfo {author} {\bibfnamefont {R.}~\bibnamefont {Osellame}}, \bibinfo {author} {\bibfnamefont {E.~F.}\ \bibnamefont {Galvão}}, \bibinfo {author} {\bibfnamefont {N.}~\bibnamefont {Spagnolo}},\ and\ \bibinfo {author} {\bibfnamefont
			{F.}~\bibnamefont {Sciarrino}},\ }\href {https://arxiv.org/abs/2407.06026} {\bibinfo {title} {Variational quantum cloning machine on a photonic integrated interferometer}} (\bibinfo {year} {2024}),\ \Eprint {https://arxiv.org/abs/2407.06026} {arXiv:2407.06026 [quant-ph]} \BibitemShut {NoStop}%
	\bibitem [{\citenamefont {Cimini}\ \emph {et~al.}(2024)\citenamefont {Cimini}, \citenamefont {Valeri}, \citenamefont {Piacentini}, \citenamefont {Ceccarelli}, \citenamefont {Corrielli}, \citenamefont {Osellame}, \citenamefont {Spagnolo},\ and\ \citenamefont {Sciarrino}}]{Cimini2024}%
	\BibitemOpen
	\bibfield  {author} {\bibinfo {author} {\bibfnamefont {V.}~\bibnamefont {Cimini}}, \bibinfo {author} {\bibfnamefont {M.}~\bibnamefont {Valeri}}, \bibinfo {author} {\bibfnamefont {S.}~\bibnamefont {Piacentini}}, \bibinfo {author} {\bibfnamefont {F.}~\bibnamefont {Ceccarelli}}, \bibinfo {author} {\bibfnamefont {G.}~\bibnamefont {Corrielli}}, \bibinfo {author} {\bibfnamefont {R.}~\bibnamefont {Osellame}}, \bibinfo {author} {\bibfnamefont {N.}~\bibnamefont {Spagnolo}},\ and\ \bibinfo {author} {\bibfnamefont {F.}~\bibnamefont {Sciarrino}},\ }\bibfield  {title} {\bibinfo {title} {Variational quantum algorithm for experimental photonic multiparameter estimation},\ }\href {https://doi.org/10.1038/s41534-024-00821-0} {\bibfield  {journal} {\bibinfo  {journal} {npj Quantum Information}\ }\textbf {\bibinfo {volume} {10}},\ \bibinfo {pages} {26} (\bibinfo {year} {2024})}\BibitemShut {NoStop}%
	\bibitem [{\citenamefont {Agresti}\ \emph {et~al.}(2024)\citenamefont {Agresti}, \citenamefont {Paul}, \citenamefont {Schiansky}, \citenamefont {Steiner}, \citenamefont {Yin}, \citenamefont {Pentangelo}, \citenamefont {Piacentini}, \citenamefont {Crespi}, \citenamefont {Ban}, \citenamefont {Ceccarelli}, \citenamefont {Osellame}, \citenamefont {Chen},\ and\ \citenamefont {Walther}}]{agresti2024}%
	\BibitemOpen
	\bibfield  {author} {\bibinfo {author} {\bibfnamefont {I.}~\bibnamefont {Agresti}}, \bibinfo {author} {\bibfnamefont {K.}~\bibnamefont {Paul}}, \bibinfo {author} {\bibfnamefont {P.}~\bibnamefont {Schiansky}}, \bibinfo {author} {\bibfnamefont {S.}~\bibnamefont {Steiner}}, \bibinfo {author} {\bibfnamefont {Z.}~\bibnamefont {Yin}}, \bibinfo {author} {\bibfnamefont {C.}~\bibnamefont {Pentangelo}}, \bibinfo {author} {\bibfnamefont {S.}~\bibnamefont {Piacentini}}, \bibinfo {author} {\bibfnamefont {A.}~\bibnamefont {Crespi}}, \bibinfo {author} {\bibfnamefont {Y.}~\bibnamefont {Ban}}, \bibinfo {author} {\bibfnamefont {F.}~\bibnamefont {Ceccarelli}}, \bibinfo {author} {\bibfnamefont {R.}~\bibnamefont {Osellame}}, \bibinfo {author} {\bibfnamefont {X.}~\bibnamefont {Chen}},\ and\ \bibinfo {author} {\bibfnamefont {P.}~\bibnamefont {Walther}},\ }\href {https://arxiv.org/abs/2408.10339} {\bibinfo {title} {Demonstration of hardware efficient photonic variational quantum algorithm}} (\bibinfo {year} {2024}),\ \Eprint
	{https://arxiv.org/abs/2408.10339} {arXiv:2408.10339 [quant-ph]} \BibitemShut {NoStop}%
	\bibitem [{\citenamefont {Zhu}\ \emph {et~al.}(2019)\citenamefont {Zhu}, \citenamefont {Linke}, \citenamefont {Benedetti}, \citenamefont {Landsman}, \citenamefont {Nguyen}, \citenamefont {Alderete}, \citenamefont {Perdomo-Ortiz}, \citenamefont {Korda}, \citenamefont {Garfoot}, \citenamefont {Brecque}, \citenamefont {Egan}, \citenamefont {Perdomo},\ and\ \citenamefont {Monroe}}]{Zhu2019}%
	\BibitemOpen
	\bibfield  {author} {\bibinfo {author} {\bibfnamefont {D.}~\bibnamefont {Zhu}}, \bibinfo {author} {\bibfnamefont {N.~M.}\ \bibnamefont {Linke}}, \bibinfo {author} {\bibfnamefont {M.}~\bibnamefont {Benedetti}}, \bibinfo {author} {\bibfnamefont {K.~A.}\ \bibnamefont {Landsman}}, \bibinfo {author} {\bibfnamefont {N.~H.}\ \bibnamefont {Nguyen}}, \bibinfo {author} {\bibfnamefont {C.~H.}\ \bibnamefont {Alderete}}, \bibinfo {author} {\bibfnamefont {A.}~\bibnamefont {Perdomo-Ortiz}}, \bibinfo {author} {\bibfnamefont {N.}~\bibnamefont {Korda}}, \bibinfo {author} {\bibfnamefont {A.}~\bibnamefont {Garfoot}}, \bibinfo {author} {\bibfnamefont {C.}~\bibnamefont {Brecque}}, \bibinfo {author} {\bibfnamefont {L.}~\bibnamefont {Egan}}, \bibinfo {author} {\bibfnamefont {O.}~\bibnamefont {Perdomo}},\ and\ \bibinfo {author} {\bibfnamefont {C.}~\bibnamefont {Monroe}},\ }\bibfield  {title} {\bibinfo {title} {Training of quantum circuits on a hybrid quantum computer},\ }\href {https://doi.org/10.1126/sciadv.aaw9918} {\bibfield
		{journal} {\bibinfo  {journal} {Science Advances}\ }\textbf {\bibinfo {volume} {5}},\ \bibinfo {pages} {eaaw9918} (\bibinfo {year} {2019})}\BibitemShut {NoStop}%
	\bibitem [{\citenamefont {Wang}\ \emph {et~al.}(2024)\citenamefont {Wang}, \citenamefont {Ding}, \citenamefont {Cárdenas-López},\ and\ \citenamefont {Chen}}]{Wang2024}%
	\BibitemOpen
	\bibfield  {author} {\bibinfo {author} {\bibfnamefont {Y.}~\bibnamefont {Wang}}, \bibinfo {author} {\bibfnamefont {Y.}~\bibnamefont {Ding}}, \bibinfo {author} {\bibfnamefont {F.~A.}\ \bibnamefont {Cárdenas-López}},\ and\ \bibinfo {author} {\bibfnamefont {X.}~\bibnamefont {Chen}},\ }\bibfield  {title} {\bibinfo {title} {Pulse-based variational quantum optimization and metalearning in superconducting circuits},\ }\href {https://doi.org/10.1103/physrevapplied.22.024009} {\bibfield  {journal} {\bibinfo  {journal} {Physical Review Applied}\ }\textbf {\bibinfo {volume} {22}},\ \bibinfo {pages} {024009} (\bibinfo {year} {2024})}\BibitemShut {NoStop}%
	\bibitem [{\citenamefont {Wang}\ \emph {et~al.}(2019)\citenamefont {Wang}, \citenamefont {Sciarrino}, \citenamefont {Laing},\ and\ \citenamefont {Thompson}}]{Wang2019}%
	\BibitemOpen
	\bibfield  {author} {\bibinfo {author} {\bibfnamefont {J.}~\bibnamefont {Wang}}, \bibinfo {author} {\bibfnamefont {F.}~\bibnamefont {Sciarrino}}, \bibinfo {author} {\bibfnamefont {A.}~\bibnamefont {Laing}},\ and\ \bibinfo {author} {\bibfnamefont {M.~G.}\ \bibnamefont {Thompson}},\ }\bibfield  {title} {\bibinfo {title} {Integrated photonic quantum technologies},\ }\href {https://doi.org/10.1038/s41566-019-0532-1} {\bibfield  {journal} {\bibinfo  {journal} {Nature Photonics}\ }\textbf {\bibinfo {volume} {14}},\ \bibinfo {pages} {273–284} (\bibinfo {year} {2019})}\BibitemShut {NoStop}%
	\bibitem [{\citenamefont {Pelucchi}\ \emph {et~al.}(2021)\citenamefont {Pelucchi}, \citenamefont {Fagas}, \citenamefont {Aharonovich}, \citenamefont {Englund}, \citenamefont {Figueroa}, \citenamefont {Gong}, \citenamefont {Hannes}, \citenamefont {Liu}, \citenamefont {Lu}, \citenamefont {Matsuda}, \citenamefont {Pan}, \citenamefont {Schreck}, \citenamefont {Sciarrino}, \citenamefont {Silberhorn}, \citenamefont {Wang},\ and\ \citenamefont {J\"{o}ns}}]{Pelucchi2021}%
	\BibitemOpen
	\bibfield  {author} {\bibinfo {author} {\bibfnamefont {E.}~\bibnamefont {Pelucchi}}, \bibinfo {author} {\bibfnamefont {G.}~\bibnamefont {Fagas}}, \bibinfo {author} {\bibfnamefont {I.}~\bibnamefont {Aharonovich}}, \bibinfo {author} {\bibfnamefont {D.}~\bibnamefont {Englund}}, \bibinfo {author} {\bibfnamefont {E.}~\bibnamefont {Figueroa}}, \bibinfo {author} {\bibfnamefont {Q.}~\bibnamefont {Gong}}, \bibinfo {author} {\bibfnamefont {H.}~\bibnamefont {Hannes}}, \bibinfo {author} {\bibfnamefont {J.}~\bibnamefont {Liu}}, \bibinfo {author} {\bibfnamefont {C.-Y.}\ \bibnamefont {Lu}}, \bibinfo {author} {\bibfnamefont {N.}~\bibnamefont {Matsuda}}, \bibinfo {author} {\bibfnamefont {J.-W.}\ \bibnamefont {Pan}}, \bibinfo {author} {\bibfnamefont {F.}~\bibnamefont {Schreck}}, \bibinfo {author} {\bibfnamefont {F.}~\bibnamefont {Sciarrino}}, \bibinfo {author} {\bibfnamefont {C.}~\bibnamefont {Silberhorn}}, \bibinfo {author} {\bibfnamefont {J.}~\bibnamefont {Wang}},\ and\ \bibinfo {author} {\bibfnamefont {K.~D.}\
			\bibnamefont {J\"{o}ns}},\ }\bibfield  {title} {\bibinfo {title} {The potential and global outlook of integrated photonics for quantum technologies},\ }\href {https://doi.org/10.1038/s42254-021-00398-z} {\bibfield  {journal} {\bibinfo  {journal} {Nature Reviews Physics}\ }\textbf {\bibinfo {volume} {4}},\ \bibinfo {pages} {194–208} (\bibinfo {year} {2021})}\BibitemShut {NoStop}%
	\bibitem [{\citenamefont {Giordani}\ \emph {et~al.}(2023{\natexlab{a}})\citenamefont {Giordani}, \citenamefont {Hoch}, \citenamefont {Carvacho}, \citenamefont {Spagnolo},\ and\ \citenamefont {Sciarrino}}]{Giordani2023}%
	\BibitemOpen
	\bibfield  {author} {\bibinfo {author} {\bibfnamefont {T.}~\bibnamefont {Giordani}}, \bibinfo {author} {\bibfnamefont {F.}~\bibnamefont {Hoch}}, \bibinfo {author} {\bibfnamefont {G.}~\bibnamefont {Carvacho}}, \bibinfo {author} {\bibfnamefont {N.}~\bibnamefont {Spagnolo}},\ and\ \bibinfo {author} {\bibfnamefont {F.}~\bibnamefont {Sciarrino}},\ }\bibfield  {title} {\bibinfo {title} {Integrated photonics in quantum technologies},\ }\href {https://doi.org/10.1007/s40766-023-00040-x} {\bibfield  {journal} {\bibinfo  {journal} {La Rivista del Nuovo Cimento}\ }\textbf {\bibinfo {volume} {46}},\ \bibinfo {pages} {71–103} (\bibinfo {year} {2023}{\natexlab{a}})}\BibitemShut {NoStop}%
	\bibitem [{\citenamefont {Paesani}\ \emph {et~al.}(2017)\citenamefont {Paesani}, \citenamefont {Gentile}, \citenamefont {Santagati}, \citenamefont {Wang}, \citenamefont {Wiebe}, \citenamefont {Tew}, \citenamefont {O'Brien},\ and\ \citenamefont {Thompson}}]{Paesani2017}%
	\BibitemOpen
	\bibfield  {author} {\bibinfo {author} {\bibfnamefont {S.}~\bibnamefont {Paesani}}, \bibinfo {author} {\bibfnamefont {A.~A.}\ \bibnamefont {Gentile}}, \bibinfo {author} {\bibfnamefont {R.}~\bibnamefont {Santagati}}, \bibinfo {author} {\bibfnamefont {J.}~\bibnamefont {Wang}}, \bibinfo {author} {\bibfnamefont {N.}~\bibnamefont {Wiebe}}, \bibinfo {author} {\bibfnamefont {D.~P.}\ \bibnamefont {Tew}}, \bibinfo {author} {\bibfnamefont {J.~L.}\ \bibnamefont {O'Brien}},\ and\ \bibinfo {author} {\bibfnamefont {M.~G.}\ \bibnamefont {Thompson}},\ }\bibfield  {title} {\bibinfo {title} {Experimental bayesian quantum phase estimation on a silicon photonic chip},\ }\href {https://doi.org/10.1103/PhysRevLett.118.100503} {\bibfield  {journal} {\bibinfo  {journal} {Phys. Rev. Lett.}\ }\textbf {\bibinfo {volume} {118}},\ \bibinfo {pages} {100503} (\bibinfo {year} {2017})}\BibitemShut {NoStop}%
	\bibitem [{\citenamefont {Jašek}\ \emph {et~al.}(2019)\citenamefont {Jašek}, \citenamefont {Jiráková}, \citenamefont {Bartkiewicz}, \citenamefont {Černoch}, \citenamefont {F\"{u}rst},\ and\ \citenamefont {Lemr}}]{Jaek2019}%
	\BibitemOpen
	\bibfield  {author} {\bibinfo {author} {\bibfnamefont {J.}~\bibnamefont {Jašek}}, \bibinfo {author} {\bibfnamefont {K.}~\bibnamefont {Jiráková}}, \bibinfo {author} {\bibfnamefont {K.}~\bibnamefont {Bartkiewicz}}, \bibinfo {author} {\bibfnamefont {A.}~\bibnamefont {Černoch}}, \bibinfo {author} {\bibfnamefont {T.}~\bibnamefont {F\"{u}rst}},\ and\ \bibinfo {author} {\bibfnamefont {K.}~\bibnamefont {Lemr}},\ }\bibfield  {title} {\bibinfo {title} {Experimental hybrid quantum-classical reinforcement learning by boson sampling: how to train a quantum cloner},\ }\href {https://doi.org/10.1364/oe.27.032454} {\bibfield  {journal} {\bibinfo  {journal} {Optics Express}\ }\textbf {\bibinfo {volume} {27}},\ \bibinfo {pages} {32454} (\bibinfo {year} {2019})}\BibitemShut {NoStop}%
	\bibitem [{\citenamefont {Schuld}\ \emph {et~al.}(2019)\citenamefont {Schuld}, \citenamefont {Bergholm}, \citenamefont {Gogolin}, \citenamefont {Izaac},\ and\ \citenamefont {Killoran}}]{schuld2019evaluating}%
	\BibitemOpen
	\bibfield  {author} {\bibinfo {author} {\bibfnamefont {M.}~\bibnamefont {Schuld}}, \bibinfo {author} {\bibfnamefont {V.}~\bibnamefont {Bergholm}}, \bibinfo {author} {\bibfnamefont {C.}~\bibnamefont {Gogolin}}, \bibinfo {author} {\bibfnamefont {J.}~\bibnamefont {Izaac}},\ and\ \bibinfo {author} {\bibfnamefont {N.}~\bibnamefont {Killoran}},\ }\bibfield  {title} {\bibinfo {title} {Evaluating analytic gradients on quantum hardware},\ }\href {https://doi.org/10.1103/PhysRevA.99.032331} {\bibfield  {journal} {\bibinfo  {journal} {Phys. Rev. A}\ }\textbf {\bibinfo {volume} {99}},\ \bibinfo {pages} {032331} (\bibinfo {year} {2019})}\BibitemShut {NoStop}%
	\bibitem [{\citenamefont {Mitarai}\ \emph {et~al.}(2018)\citenamefont {Mitarai}, \citenamefont {Negoro}, \citenamefont {Kitagawa},\ and\ \citenamefont {Fujii}}]{mitarai2018quantum}%
	\BibitemOpen
	\bibfield  {author} {\bibinfo {author} {\bibfnamefont {K.}~\bibnamefont {Mitarai}}, \bibinfo {author} {\bibfnamefont {M.}~\bibnamefont {Negoro}}, \bibinfo {author} {\bibfnamefont {M.}~\bibnamefont {Kitagawa}},\ and\ \bibinfo {author} {\bibfnamefont {K.}~\bibnamefont {Fujii}},\ }\bibfield  {title} {\bibinfo {title} {Quantum circuit learning},\ }\href {https://doi.org/10.1103/PhysRevA.98.032309} {\bibfield  {journal} {\bibinfo  {journal} {Phys. Rev. A}\ }\textbf {\bibinfo {volume} {98}},\ \bibinfo {pages} {032309} (\bibinfo {year} {2018})}\BibitemShut {NoStop}%
	\bibitem [{\citenamefont {Wierichs}\ \emph {et~al.}(2022)\citenamefont {Wierichs}, \citenamefont {Izaac}, \citenamefont {Wang},\ and\ \citenamefont {Lin}}]{Wierichs2022}%
	\BibitemOpen
	\bibfield  {author} {\bibinfo {author} {\bibfnamefont {D.}~\bibnamefont {Wierichs}}, \bibinfo {author} {\bibfnamefont {J.}~\bibnamefont {Izaac}}, \bibinfo {author} {\bibfnamefont {C.}~\bibnamefont {Wang}},\ and\ \bibinfo {author} {\bibfnamefont {C.~Y.-Y.}\ \bibnamefont {Lin}},\ }\bibfield  {title} {\bibinfo {title} {General parameter-shift rules for quantum gradients},\ }\href {https://doi.org/10.22331/q-2022-03-30-677} {\bibfield  {journal} {\bibinfo  {journal} {Quantum}\ }\textbf {\bibinfo {volume} {6}},\ \bibinfo {pages} {677} (\bibinfo {year} {2022})}\BibitemShut {NoStop}%
	\bibitem [{\citenamefont {Gan}\ \emph {et~al.}(2022)\citenamefont {Gan}, \citenamefont {Leykam},\ and\ \citenamefont {Angelakis}}]{Gan2022}%
	\BibitemOpen
	\bibfield  {author} {\bibinfo {author} {\bibfnamefont {B.~Y.}\ \bibnamefont {Gan}}, \bibinfo {author} {\bibfnamefont {D.}~\bibnamefont {Leykam}},\ and\ \bibinfo {author} {\bibfnamefont {D.~G.}\ \bibnamefont {Angelakis}},\ }\bibfield  {title} {\bibinfo {title} {Fock state-enhanced expressivity of quantum machine learning models},\ }\bibfield  {journal} {\bibinfo  {journal} {EPJ Quantum Technology}\ }\textbf {\bibinfo {volume} {9}},\ \href {https://doi.org/10.1140/epjqt/s40507-022-00135-0} {10.1140/epjqt/s40507-022-00135-0} (\bibinfo {year} {2022})\BibitemShut {NoStop}%
	\bibitem [{\citenamefont {Aaronson}\ and\ \citenamefont {Arkhipov}(2013)}]{Aaronson2013}%
	\BibitemOpen
	\bibfield  {author} {\bibinfo {author} {\bibfnamefont {S.}~\bibnamefont {Aaronson}}\ and\ \bibinfo {author} {\bibfnamefont {A.}~\bibnamefont {Arkhipov}},\ }\bibfield  {title} {\bibinfo {title} {The computational complexity of linear optics},\ }\href {https://doi.org/10.4086/toc.2013.v009a004} {\bibfield  {journal} {\bibinfo  {journal} {Theory of Computing}\ }\textbf {\bibinfo {volume} {9}},\ \bibinfo {pages} {143–252} (\bibinfo {year} {2013})}\BibitemShut {NoStop}%
	\bibitem [{\citenamefont {Sakurai}(1993)}]{Sakurai1993}%
	\BibitemOpen
	\bibfield  {author} {\bibinfo {author} {\bibfnamefont {J.~J.}\ \bibnamefont {Sakurai}},\ }\href@noop {} {\emph {\bibinfo {title} {Modern Quantum Mechanics, Revised Edition}}}\ (\bibinfo  {publisher} {Pearson},\ \bibinfo {address} {Upper Saddle River, NJ},\ \bibinfo {year} {1993})\BibitemShut {NoStop}%
	\bibitem [{\citenamefont {Buzcircek}\ \emph {et~al.}(2000)\citenamefont {Buzcircek}, \citenamefont {Hillery},\ and\ \citenamefont {Werner}}]{Buzcircek2000}%
	\BibitemOpen
	\bibfield  {author} {\bibinfo {author} {\bibfnamefont {V.}~\bibnamefont {Buzcircek}}, \bibinfo {author} {\bibfnamefont {M.}~\bibnamefont {Hillery}},\ and\ \bibinfo {author} {\bibfnamefont {F.}~\bibnamefont {Werner}},\ }\bibfield  {title} {\bibinfo {title} {Universal-not gate},\ }\href {https://doi.org/10.1080/09500340008244037} {\bibfield  {journal} {\bibinfo  {journal} {Journal of Modern Optics}\ }\textbf {\bibinfo {volume} {47}},\ \bibinfo {pages} {211–232} (\bibinfo {year} {2000})}\BibitemShut {NoStop}%
	\bibitem [{\citenamefont {De~Martini}\ \emph {et~al.}(2002)\citenamefont {De~Martini}, \citenamefont {Bužek}, \citenamefont {Sciarrino},\ and\ \citenamefont {Sias}}]{DeMartini2002}%
	\BibitemOpen
	\bibfield  {author} {\bibinfo {author} {\bibfnamefont {F.}~\bibnamefont {De~Martini}}, \bibinfo {author} {\bibfnamefont {V.}~\bibnamefont {Bužek}}, \bibinfo {author} {\bibfnamefont {F.}~\bibnamefont {Sciarrino}},\ and\ \bibinfo {author} {\bibfnamefont {C.}~\bibnamefont {Sias}},\ }\bibfield  {title} {\bibinfo {title} {Experimental realization of the quantum universal not gate},\ }\href {https://doi.org/10.1038/nature01093} {\bibfield  {journal} {\bibinfo  {journal} {Nature}\ }\textbf {\bibinfo {volume} {419}},\ \bibinfo {pages} {815–818} (\bibinfo {year} {2002})}\BibitemShut {NoStop}%
	\bibitem [{\citenamefont {Ricci}\ \emph {et~al.}(2004)\citenamefont {Ricci}, \citenamefont {Sciarrino}, \citenamefont {Sias},\ and\ \citenamefont {De~Martini}}]{Ricci2004}%
	\BibitemOpen
	\bibfield  {author} {\bibinfo {author} {\bibfnamefont {M.}~\bibnamefont {Ricci}}, \bibinfo {author} {\bibfnamefont {F.}~\bibnamefont {Sciarrino}}, \bibinfo {author} {\bibfnamefont {C.}~\bibnamefont {Sias}},\ and\ \bibinfo {author} {\bibfnamefont {F.}~\bibnamefont {De~Martini}},\ }\bibfield  {title} {\bibinfo {title} {Teleportation scheme implementing the universal optimal quantum cloning machine and the universal not gate},\ }\bibfield  {journal} {\bibinfo  {journal} {Physical Review Letters}\ }\textbf {\bibinfo {volume} {92}},\ \href {https://doi.org/10.1103/physrevlett.92.047901} {10.1103/physrevlett.92.047901} (\bibinfo {year} {2004})\BibitemShut {NoStop}%
	\bibitem [{\citenamefont {Zygmund}(2015)}]{trigonometric2015}%
	\BibitemOpen
	\bibfield  {author} {\bibinfo {author} {\bibfnamefont {A.}~\bibnamefont {Zygmund}},\ }\href@noop {} {\emph {\bibinfo {title} {Trigonometric series}}},\ \bibinfo {edition} {3rd}\ ed.\ (\bibinfo  {publisher} {Cambridge University Press},\ \bibinfo {address} {Cambridge, England},\ \bibinfo {year} {2015})\BibitemShut {NoStop}%
	\bibitem [{\citenamefont {Kok}\ \emph {et~al.}(2007)\citenamefont {Kok}, \citenamefont {Munro}, \citenamefont {Nemoto}, \citenamefont {Ralph}, \citenamefont {Dowling},\ and\ \citenamefont {Milburn}}]{kok2007linear}%
	\BibitemOpen
	\bibfield  {author} {\bibinfo {author} {\bibfnamefont {P.}~\bibnamefont {Kok}}, \bibinfo {author} {\bibfnamefont {W.~J.}\ \bibnamefont {Munro}}, \bibinfo {author} {\bibfnamefont {K.}~\bibnamefont {Nemoto}}, \bibinfo {author} {\bibfnamefont {T.~C.}\ \bibnamefont {Ralph}}, \bibinfo {author} {\bibfnamefont {J.~P.}\ \bibnamefont {Dowling}},\ and\ \bibinfo {author} {\bibfnamefont {G.~J.}\ \bibnamefont {Milburn}},\ }\bibfield  {title} {\bibinfo {title} {Linear optical quantum computing with photonic qubits},\ }\href@noop {} {\bibfield  {journal} {\bibinfo  {journal} {Reviews of modern physics}\ }\textbf {\bibinfo {volume} {79}},\ \bibinfo {pages} {135} (\bibinfo {year} {2007})}\BibitemShut {NoStop}%
	\bibitem [{\citenamefont {Clements}\ \emph {et~al.}(2016)\citenamefont {Clements}, \citenamefont {Humphreys}, \citenamefont {Metcalf}, \citenamefont {Kolthammer},\ and\ \citenamefont {Walsmley}}]{Clements2016}%
	\BibitemOpen
	\bibfield  {author} {\bibinfo {author} {\bibfnamefont {W.~R.}\ \bibnamefont {Clements}}, \bibinfo {author} {\bibfnamefont {P.~C.}\ \bibnamefont {Humphreys}}, \bibinfo {author} {\bibfnamefont {B.~J.}\ \bibnamefont {Metcalf}}, \bibinfo {author} {\bibfnamefont {W.~S.}\ \bibnamefont {Kolthammer}},\ and\ \bibinfo {author} {\bibfnamefont {I.~A.}\ \bibnamefont {Walsmley}},\ }\bibfield  {title} {\bibinfo {title} {Optimal design for universal multiport interferometers},\ }\href {https://doi.org/10.1364/optica.3.001460} {\bibfield  {journal} {\bibinfo  {journal} {Optica}\ }\textbf {\bibinfo {volume} {3}},\ \bibinfo {pages} {1460} (\bibinfo {year} {2016})}\BibitemShut {NoStop}%
	\bibitem [{\citenamefont {Tichy}\ \emph {et~al.}(2012)\citenamefont {Tichy}, \citenamefont {Tiersch}, \citenamefont {Mintert},\ and\ \citenamefont {Buchleitner}}]{Tichy2012}%
	\BibitemOpen
	\bibfield  {author} {\bibinfo {author} {\bibfnamefont {M.~C.}\ \bibnamefont {Tichy}}, \bibinfo {author} {\bibfnamefont {M.}~\bibnamefont {Tiersch}}, \bibinfo {author} {\bibfnamefont {F.}~\bibnamefont {Mintert}},\ and\ \bibinfo {author} {\bibfnamefont {A.}~\bibnamefont {Buchleitner}},\ }\bibfield  {title} {\bibinfo {title} {Many-particle interference beyond many-boson and many-fermion statistics},\ }\href {https://doi.org/10.1088/1367-2630/14/9/093015} {\bibfield  {journal} {\bibinfo  {journal} {New Journal of Physics}\ }\textbf {\bibinfo {volume} {14}},\ \bibinfo {pages} {093015} (\bibinfo {year} {2012})}\BibitemShut {NoStop}%
	\bibitem [{\citenamefont {Brod}\ \emph {et~al.}(2019)\citenamefont {Brod}, \citenamefont {Galv{\~a}o}, \citenamefont {Crespi}, \citenamefont {Osellame}, \citenamefont {Spagnolo},\ and\ \citenamefont {Sciarrino}}]{brod2019photonic}%
	\BibitemOpen
	\bibfield  {author} {\bibinfo {author} {\bibfnamefont {D.~J.}\ \bibnamefont {Brod}}, \bibinfo {author} {\bibfnamefont {E.~F.}\ \bibnamefont {Galv{\~a}o}}, \bibinfo {author} {\bibfnamefont {A.}~\bibnamefont {Crespi}}, \bibinfo {author} {\bibfnamefont {R.}~\bibnamefont {Osellame}}, \bibinfo {author} {\bibfnamefont {N.}~\bibnamefont {Spagnolo}},\ and\ \bibinfo {author} {\bibfnamefont {F.}~\bibnamefont {Sciarrino}},\ }\bibfield  {title} {\bibinfo {title} {Photonic implementation of boson sampling: a review},\ }\href@noop {} {\bibfield  {journal} {\bibinfo  {journal} {Advanced Photonics}\ }\textbf {\bibinfo {volume} {1}},\ \bibinfo {pages} {034001} (\bibinfo {year} {2019})}\BibitemShut {NoStop}%
	\bibitem [{\citenamefont {Shchesnovich}(2015)}]{Shchesnovich2015}%
	\BibitemOpen
	\bibfield  {author} {\bibinfo {author} {\bibfnamefont {V.~S.}\ \bibnamefont {Shchesnovich}},\ }\bibfield  {title} {\bibinfo {title} {Partial indistinguishability theory for multiphoton experiments in multiport devices},\ }\href {https://doi.org/10.1103/physreva.91.013844} {\bibfield  {journal} {\bibinfo  {journal} {Physical Review A}\ }\textbf {\bibinfo {volume} {91}},\ \bibinfo {pages} {013844} (\bibinfo {year} {2015})}\BibitemShut {NoStop}%
	\bibitem [{\citenamefont {Coyle}\ \emph {et~al.}(2022)\citenamefont {Coyle}, \citenamefont {Doosti}, \citenamefont {Kashefi},\ and\ \citenamefont {Kumar}}]{coyle2022progress}%
	\BibitemOpen
	\bibfield  {author} {\bibinfo {author} {\bibfnamefont {B.}~\bibnamefont {Coyle}}, \bibinfo {author} {\bibfnamefont {M.}~\bibnamefont {Doosti}}, \bibinfo {author} {\bibfnamefont {E.}~\bibnamefont {Kashefi}},\ and\ \bibinfo {author} {\bibfnamefont {N.}~\bibnamefont {Kumar}},\ }\bibfield  {title} {\bibinfo {title} {Progress toward practical quantum cryptanalysis by variational quantum cloning},\ }\href@noop {} {\bibfield  {journal} {\bibinfo  {journal} {Physical Review A}\ }\textbf {\bibinfo {volume} {105}},\ \bibinfo {pages} {042604} (\bibinfo {year} {2022})}\BibitemShut {NoStop}%
	\bibitem [{\citenamefont {Corrielli}\ \emph {et~al.}(2021)\citenamefont {Corrielli}, \citenamefont {Crespi},\ and\ \citenamefont {Osellame}}]{Corrielli2021}%
	\BibitemOpen
	\bibfield  {author} {\bibinfo {author} {\bibfnamefont {G.}~\bibnamefont {Corrielli}}, \bibinfo {author} {\bibfnamefont {A.}~\bibnamefont {Crespi}},\ and\ \bibinfo {author} {\bibfnamefont {R.}~\bibnamefont {Osellame}},\ }\bibfield  {title} {\bibinfo {title} {Femtosecond laser micromachining for integrated quantum photonics},\ }\href {https://doi.org/10.1515/nanoph-2021-0419} {\bibfield  {journal} {\bibinfo  {journal} {Nanophotonics}\ }\textbf {\bibinfo {volume} {10}},\ \bibinfo {pages} {3789–3812} (\bibinfo {year} {2021})}\BibitemShut {NoStop}%
	\bibitem [{\citenamefont {Pentangelo}\ \emph {et~al.}(2024)\citenamefont {Pentangelo}, \citenamefont {Di~Giano}, \citenamefont {Piacentini}, \citenamefont {Arpe}, \citenamefont {Ceccarelli}, \citenamefont {Crespi},\ and\ \citenamefont {Osellame}}]{Pentangelo2024}%
	\BibitemOpen
	\bibfield  {author} {\bibinfo {author} {\bibfnamefont {C.}~\bibnamefont {Pentangelo}}, \bibinfo {author} {\bibfnamefont {N.}~\bibnamefont {Di~Giano}}, \bibinfo {author} {\bibfnamefont {S.}~\bibnamefont {Piacentini}}, \bibinfo {author} {\bibfnamefont {R.}~\bibnamefont {Arpe}}, \bibinfo {author} {\bibfnamefont {F.}~\bibnamefont {Ceccarelli}}, \bibinfo {author} {\bibfnamefont {A.}~\bibnamefont {Crespi}},\ and\ \bibinfo {author} {\bibfnamefont {R.}~\bibnamefont {Osellame}},\ }\bibfield  {title} {\bibinfo {title} {High-fidelity and polarization-insensitive universal photonic processors fabricated by femtosecond laser writing},\ }\href {https://doi.org/10.1515/nanoph-2023-0636} {\bibfield  {journal} {\bibinfo  {journal} {Nanophotonics}\ }\textbf {\bibinfo {volume} {13}},\ \bibinfo {pages} {2259–2270} (\bibinfo {year} {2024})}\BibitemShut {NoStop}%
	\bibitem [{\citenamefont {Giordani}\ \emph {et~al.}(2023{\natexlab{b}})\citenamefont {Giordani}, \citenamefont {Wagner}, \citenamefont {Esposito}, \citenamefont {Camillini}, \citenamefont {Hoch}, \citenamefont {Carvacho}, \citenamefont {Pentangelo}, \citenamefont {Ceccarelli}, \citenamefont {Piacentini}, \citenamefont {Crespi}, \citenamefont {Spagnolo}, \citenamefont {Osellame}, \citenamefont {Galvão},\ and\ \citenamefont {Sciarrino}}]{Giordani24}%
	\BibitemOpen
	\bibfield  {author} {\bibinfo {author} {\bibfnamefont {T.}~\bibnamefont {Giordani}}, \bibinfo {author} {\bibfnamefont {R.}~\bibnamefont {Wagner}}, \bibinfo {author} {\bibfnamefont {C.}~\bibnamefont {Esposito}}, \bibinfo {author} {\bibfnamefont {A.}~\bibnamefont {Camillini}}, \bibinfo {author} {\bibfnamefont {F.}~\bibnamefont {Hoch}}, \bibinfo {author} {\bibfnamefont {G.}~\bibnamefont {Carvacho}}, \bibinfo {author} {\bibfnamefont {C.}~\bibnamefont {Pentangelo}}, \bibinfo {author} {\bibfnamefont {F.}~\bibnamefont {Ceccarelli}}, \bibinfo {author} {\bibfnamefont {S.}~\bibnamefont {Piacentini}}, \bibinfo {author} {\bibfnamefont {A.}~\bibnamefont {Crespi}}, \bibinfo {author} {\bibfnamefont {N.}~\bibnamefont {Spagnolo}}, \bibinfo {author} {\bibfnamefont {R.}~\bibnamefont {Osellame}}, \bibinfo {author} {\bibfnamefont {E.~F.}\ \bibnamefont {Galvão}},\ and\ \bibinfo {author} {\bibfnamefont {F.}~\bibnamefont {Sciarrino}},\ }\bibfield  {title} {\bibinfo {title} {Experimental certification of contextuality, coherence,
			and dimension in a programmable universal photonic processor},\ }\href {https://doi.org/10.1126/sciadv.adj4249} {\bibfield  {journal} {\bibinfo  {journal} {Science Advances}\ }\textbf {\bibinfo {volume} {9}},\ \bibinfo {pages} {eadj4249} (\bibinfo {year} {2023}{\natexlab{b}})}\BibitemShut {NoStop}%
	\bibitem [{\citenamefont {McClean}\ \emph {et~al.}(2020)\citenamefont {McClean}, \citenamefont {Rubin}, \citenamefont {Sung}, \citenamefont {Kivlichan}, \citenamefont {Bonet-Monroig}, \citenamefont {Cao}, \citenamefont {Dai}, \citenamefont {Fried}, \citenamefont {Gidney}, \citenamefont {Gimby}, \citenamefont {Gokhale}, \citenamefont {H\"{a}ner}, \citenamefont {Hardikar}, \citenamefont {Havlíček}, \citenamefont {Higgott}, \citenamefont {Huang}, \citenamefont {Izaac}, \citenamefont {Jiang}, \citenamefont {Liu}, \citenamefont {McArdle}, \citenamefont {Neeley}, \citenamefont {O’Brien}, \citenamefont {O’Gorman}, \citenamefont {Ozfidan}, \citenamefont {Radin}, \citenamefont {Romero}, \citenamefont {Sawaya}, \citenamefont {Senjean}, \citenamefont {Setia}, \citenamefont {Sim}, \citenamefont {Steiger}, \citenamefont {Steudtner}, \citenamefont {Sun}, \citenamefont {Sun}, \citenamefont {Wang}, \citenamefont {Zhang},\ and\ \citenamefont {Babbush}}]{McClean2020}%
	\BibitemOpen
	\bibfield  {author} {\bibinfo {author} {\bibfnamefont {J.~R.}\ \bibnamefont {McClean}}, \bibinfo {author} {\bibfnamefont {N.~C.}\ \bibnamefont {Rubin}}, \bibinfo {author} {\bibfnamefont {K.~J.}\ \bibnamefont {Sung}}, \bibinfo {author} {\bibfnamefont {I.~D.}\ \bibnamefont {Kivlichan}}, \bibinfo {author} {\bibfnamefont {X.}~\bibnamefont {Bonet-Monroig}}, \bibinfo {author} {\bibfnamefont {Y.}~\bibnamefont {Cao}}, \bibinfo {author} {\bibfnamefont {C.}~\bibnamefont {Dai}}, \bibinfo {author} {\bibfnamefont {E.~S.}\ \bibnamefont {Fried}}, \bibinfo {author} {\bibfnamefont {C.}~\bibnamefont {Gidney}}, \bibinfo {author} {\bibfnamefont {B.}~\bibnamefont {Gimby}}, \bibinfo {author} {\bibfnamefont {P.}~\bibnamefont {Gokhale}}, \bibinfo {author} {\bibfnamefont {T.}~\bibnamefont {H\"{a}ner}}, \bibinfo {author} {\bibfnamefont {T.}~\bibnamefont {Hardikar}}, \bibinfo {author} {\bibfnamefont {V.}~\bibnamefont {Havlíček}}, \bibinfo {author} {\bibfnamefont {O.}~\bibnamefont {Higgott}}, \bibinfo {author} {\bibfnamefont
			{C.}~\bibnamefont {Huang}}, \bibinfo {author} {\bibfnamefont {J.}~\bibnamefont {Izaac}}, \bibinfo {author} {\bibfnamefont {Z.}~\bibnamefont {Jiang}}, \bibinfo {author} {\bibfnamefont {X.}~\bibnamefont {Liu}}, \bibinfo {author} {\bibfnamefont {S.}~\bibnamefont {McArdle}}, \bibinfo {author} {\bibfnamefont {M.}~\bibnamefont {Neeley}}, \bibinfo {author} {\bibfnamefont {T.}~\bibnamefont {O’Brien}}, \bibinfo {author} {\bibfnamefont {B.}~\bibnamefont {O’Gorman}}, \bibinfo {author} {\bibfnamefont {I.}~\bibnamefont {Ozfidan}}, \bibinfo {author} {\bibfnamefont {M.~D.}\ \bibnamefont {Radin}}, \bibinfo {author} {\bibfnamefont {J.}~\bibnamefont {Romero}}, \bibinfo {author} {\bibfnamefont {N.~P.~D.}\ \bibnamefont {Sawaya}}, \bibinfo {author} {\bibfnamefont {B.}~\bibnamefont {Senjean}}, \bibinfo {author} {\bibfnamefont {K.}~\bibnamefont {Setia}}, \bibinfo {author} {\bibfnamefont {S.}~\bibnamefont {Sim}}, \bibinfo {author} {\bibfnamefont {D.~S.}\ \bibnamefont {Steiger}}, \bibinfo {author} {\bibfnamefont
			{M.}~\bibnamefont {Steudtner}}, \bibinfo {author} {\bibfnamefont {Q.}~\bibnamefont {Sun}}, \bibinfo {author} {\bibfnamefont {W.}~\bibnamefont {Sun}}, \bibinfo {author} {\bibfnamefont {D.}~\bibnamefont {Wang}}, \bibinfo {author} {\bibfnamefont {F.}~\bibnamefont {Zhang}},\ and\ \bibinfo {author} {\bibfnamefont {R.}~\bibnamefont {Babbush}},\ }\bibfield  {title} {\bibinfo {title} {Openfermion: the electronic structure package for quantum computers},\ }\href {https://doi.org/10.1088/2058-9565/ab8ebc} {\bibfield  {journal} {\bibinfo  {journal} {Quantum Science and Technology}\ }\textbf {\bibinfo {volume} {5}},\ \bibinfo {pages} {034014} (\bibinfo {year} {2020})}\BibitemShut {NoStop}%
	\bibitem [{\citenamefont {Nelder}\ and\ \citenamefont {Mead}(1965)}]{Nelder1965}%
	\BibitemOpen
	\bibfield  {author} {\bibinfo {author} {\bibfnamefont {J.~A.}\ \bibnamefont {Nelder}}\ and\ \bibinfo {author} {\bibfnamefont {R.}~\bibnamefont {Mead}},\ }\bibfield  {title} {\bibinfo {title} {A simplex method for function minimization},\ }\href {https://doi.org/10.1093/comjnl/7.4.308} {\bibfield  {journal} {\bibinfo  {journal} {The Computer Journal}\ }\textbf {\bibinfo {volume} {7}},\ \bibinfo {pages} {308–313} (\bibinfo {year} {1965})}\BibitemShut {NoStop}%
	\bibitem [{\citenamefont {Fletcher}(2000)}]{Fletcher2000}%
	\BibitemOpen
	\bibfield  {author} {\bibinfo {author} {\bibfnamefont {R.}~\bibnamefont {Fletcher}},\ }\href {https://doi.org/10.1002/9781118723203} {\emph {\bibinfo {title} {Practical Methods of Optimization}}}\ (\bibinfo  {publisher} {Wiley},\ \bibinfo {year} {2000})\BibitemShut {NoStop}%
	\bibitem [{\citenamefont {Virtanen}\ \emph {et~al.}(2020)\citenamefont {Virtanen}, \citenamefont {Gommers}, \citenamefont {Oliphant}, \citenamefont {Haberland}, \citenamefont {Reddy}, \citenamefont {Cournapeau}, \citenamefont {Burovski}, \citenamefont {Peterson}, \citenamefont {Weckesser}, \citenamefont {Bright}, \citenamefont {{van der Walt}}, \citenamefont {Brett}, \citenamefont {Wilson}, \citenamefont {Millman}, \citenamefont {Mayorov}, \citenamefont {Nelson}, \citenamefont {Jones}, \citenamefont {Kern}, \citenamefont {Larson}, \citenamefont {Carey}, \citenamefont {Polat}, \citenamefont {Feng}, \citenamefont {Moore}, \citenamefont {{VanderPlas}}, \citenamefont {Laxalde}, \citenamefont {Perktold}, \citenamefont {Cimrman}, \citenamefont {Henriksen}, \citenamefont {Quintero}, \citenamefont {Harris}, \citenamefont {Archibald}, \citenamefont {Ribeiro}, \citenamefont {Pedregosa}, \citenamefont {{van Mulbregt}},\ and\ \citenamefont {{SciPy 1.0 Contributors}}}]{2020SciPy-NMeth}%
	\BibitemOpen
	\bibfield  {author} {\bibinfo {author} {\bibfnamefont {P.}~\bibnamefont {Virtanen}}, \bibinfo {author} {\bibfnamefont {R.}~\bibnamefont {Gommers}}, \bibinfo {author} {\bibfnamefont {T.~E.}\ \bibnamefont {Oliphant}}, \bibinfo {author} {\bibfnamefont {M.}~\bibnamefont {Haberland}}, \bibinfo {author} {\bibfnamefont {T.}~\bibnamefont {Reddy}}, \bibinfo {author} {\bibfnamefont {D.}~\bibnamefont {Cournapeau}}, \bibinfo {author} {\bibfnamefont {E.}~\bibnamefont {Burovski}}, \bibinfo {author} {\bibfnamefont {P.}~\bibnamefont {Peterson}}, \bibinfo {author} {\bibfnamefont {W.}~\bibnamefont {Weckesser}}, \bibinfo {author} {\bibfnamefont {J.}~\bibnamefont {Bright}}, \bibinfo {author} {\bibfnamefont {S.~J.}\ \bibnamefont {{van der Walt}}}, \bibinfo {author} {\bibfnamefont {M.}~\bibnamefont {Brett}}, \bibinfo {author} {\bibfnamefont {J.}~\bibnamefont {Wilson}}, \bibinfo {author} {\bibfnamefont {K.~J.}\ \bibnamefont {Millman}}, \bibinfo {author} {\bibfnamefont {N.}~\bibnamefont {Mayorov}}, \bibinfo {author} {\bibfnamefont
			{A.~R.~J.}\ \bibnamefont {Nelson}}, \bibinfo {author} {\bibfnamefont {E.}~\bibnamefont {Jones}}, \bibinfo {author} {\bibfnamefont {R.}~\bibnamefont {Kern}}, \bibinfo {author} {\bibfnamefont {E.}~\bibnamefont {Larson}}, \bibinfo {author} {\bibfnamefont {C.~J.}\ \bibnamefont {Carey}}, \bibinfo {author} {\bibfnamefont {{\.I}.}~\bibnamefont {Polat}}, \bibinfo {author} {\bibfnamefont {Y.}~\bibnamefont {Feng}}, \bibinfo {author} {\bibfnamefont {E.~W.}\ \bibnamefont {Moore}}, \bibinfo {author} {\bibfnamefont {J.}~\bibnamefont {{VanderPlas}}}, \bibinfo {author} {\bibfnamefont {D.}~\bibnamefont {Laxalde}}, \bibinfo {author} {\bibfnamefont {J.}~\bibnamefont {Perktold}}, \bibinfo {author} {\bibfnamefont {R.}~\bibnamefont {Cimrman}}, \bibinfo {author} {\bibfnamefont {I.}~\bibnamefont {Henriksen}}, \bibinfo {author} {\bibfnamefont {E.~A.}\ \bibnamefont {Quintero}}, \bibinfo {author} {\bibfnamefont {C.~R.}\ \bibnamefont {Harris}}, \bibinfo {author} {\bibfnamefont {A.~M.}\ \bibnamefont {Archibald}}, \bibinfo {author}
		{\bibfnamefont {A.~H.}\ \bibnamefont {Ribeiro}}, \bibinfo {author} {\bibfnamefont {F.}~\bibnamefont {Pedregosa}}, \bibinfo {author} {\bibfnamefont {P.}~\bibnamefont {{van Mulbregt}}},\ and\ \bibinfo {author} {\bibnamefont {{SciPy 1.0 Contributors}}},\ }\bibfield  {title} {\bibinfo {title} {{{SciPy} 1.0: Fundamental Algorithms for Scientific Computing in Python}},\ }\href {https://doi.org/10.1038/s41592-019-0686-2} {\bibfield  {journal} {\bibinfo  {journal} {Nature Methods}\ }\textbf {\bibinfo {volume} {17}},\ \bibinfo {pages} {261} (\bibinfo {year} {2020})}\BibitemShut {NoStop}%
	\bibitem [{\citenamefont {Facelli}\ \emph {et~al.}(2024)\citenamefont {Facelli}, \citenamefont {Roberts}, \citenamefont {Wallner}, \citenamefont {Makarovskiy}, \citenamefont {Holmes},\ and\ \citenamefont {Clements}}]{facelli2024exactgradientslinearoptics}%
	\BibitemOpen
	\bibfield  {author} {\bibinfo {author} {\bibfnamefont {G.}~\bibnamefont {Facelli}}, \bibinfo {author} {\bibfnamefont {D.~D.}\ \bibnamefont {Roberts}}, \bibinfo {author} {\bibfnamefont {H.}~\bibnamefont {Wallner}}, \bibinfo {author} {\bibfnamefont {A.}~\bibnamefont {Makarovskiy}}, \bibinfo {author} {\bibfnamefont {Z.}~\bibnamefont {Holmes}},\ and\ \bibinfo {author} {\bibfnamefont {W.~R.}\ \bibnamefont {Clements}},\ }\href {https://arxiv.org/abs/2409.16369} {\bibinfo {title} {Exact gradients for linear optics with single photons}} (\bibinfo {year} {2024}),\ \Eprint {https://arxiv.org/abs/2409.16369} {arXiv:2409.16369 [quant-ph]} \BibitemShut {NoStop}%
	\bibitem [{\citenamefont {Pappalardo}\ \emph {et~al.}(2024)\citenamefont {Pappalardo}, \citenamefont {Emeriau}, \citenamefont {de~Felice}, \citenamefont {Ventura}, \citenamefont {Jaunin}, \citenamefont {Yeung}, \citenamefont {Coecke},\ and\ \citenamefont {Mansfield}}]{pappalardo2024photonicparametershiftruleenabling}%
	\BibitemOpen
	\bibfield  {author} {\bibinfo {author} {\bibfnamefont {A.}~\bibnamefont {Pappalardo}}, \bibinfo {author} {\bibfnamefont {P.-E.}\ \bibnamefont {Emeriau}}, \bibinfo {author} {\bibfnamefont {G.}~\bibnamefont {de~Felice}}, \bibinfo {author} {\bibfnamefont {B.}~\bibnamefont {Ventura}}, \bibinfo {author} {\bibfnamefont {H.}~\bibnamefont {Jaunin}}, \bibinfo {author} {\bibfnamefont {R.}~\bibnamefont {Yeung}}, \bibinfo {author} {\bibfnamefont {B.}~\bibnamefont {Coecke}},\ and\ \bibinfo {author} {\bibfnamefont {S.}~\bibnamefont {Mansfield}},\ }\href {https://arxiv.org/abs/2410.02726} {\bibinfo {title} {A photonic parameter-shift rule: Enabling gradient computation for photonic quantum computers}} (\bibinfo {year} {2024}),\ \Eprint {https://arxiv.org/abs/2410.02726} {arXiv:2410.02726 [quant-ph]} \BibitemShut {NoStop}%
	\bibitem [{\citenamefont {de~Felice}\ and\ \citenamefont {Corlett}(2024)}]{defelice2024}%
	\BibitemOpen
	\bibfield  {author} {\bibinfo {author} {\bibfnamefont {G.}~\bibnamefont {de~Felice}}\ and\ \bibinfo {author} {\bibfnamefont {C.}~\bibnamefont {Corlett}},\ }\href {https://arxiv.org/abs/2401.07997} {\bibinfo {title} {Differentiation of linear optical circuits}} (\bibinfo {year} {2024}),\ \Eprint {https://arxiv.org/abs/2401.07997} {arXiv:2401.07997 [quant-ph]} \BibitemShut {NoStop}%
	\bibitem [{\citenamefont {Bowles}\ \emph {et~al.}(2024)\citenamefont {Bowles}, \citenamefont {Wierichs},\ and\ \citenamefont {Park}}]{bowles2024}%
	\BibitemOpen
	\bibfield  {author} {\bibinfo {author} {\bibfnamefont {J.}~\bibnamefont {Bowles}}, \bibinfo {author} {\bibfnamefont {D.}~\bibnamefont {Wierichs}},\ and\ \bibinfo {author} {\bibfnamefont {C.-Y.}\ \bibnamefont {Park}},\ }\href {https://arxiv.org/abs/2306.14962} {\bibinfo {title} {Backpropagation scaling in parameterised quantum circuits}} (\bibinfo {year} {2024}),\ \Eprint {https://arxiv.org/abs/2306.14962} {arXiv:2306.14962 [quant-ph]} \BibitemShut {NoStop}%
	\bibitem [{\citenamefont {Abbas}\ \emph {et~al.}(2023)\citenamefont {Abbas}, \citenamefont {King}, \citenamefont {Huang}, \citenamefont {Huggins}, \citenamefont {Movassagh}, \citenamefont {Gilboa},\ and\ \citenamefont {McClean}}]{abbas2023}%
	\BibitemOpen
	\bibfield  {author} {\bibinfo {author} {\bibfnamefont {A.}~\bibnamefont {Abbas}}, \bibinfo {author} {\bibfnamefont {R.}~\bibnamefont {King}}, \bibinfo {author} {\bibfnamefont {H.-Y.}\ \bibnamefont {Huang}}, \bibinfo {author} {\bibfnamefont {W.~J.}\ \bibnamefont {Huggins}}, \bibinfo {author} {\bibfnamefont {R.}~\bibnamefont {Movassagh}}, \bibinfo {author} {\bibfnamefont {D.}~\bibnamefont {Gilboa}},\ and\ \bibinfo {author} {\bibfnamefont {J.~R.}\ \bibnamefont {McClean}},\ }\bibfield  {title} {\bibinfo {title} {On quantum backpropagation, information reuse, and cheating measurement collapse},\ }in\ \href {https://openreview.net/forum?id=HF6bnhfSqH} {\emph {\bibinfo {booktitle} {Thirty-seventh Conference on Neural Information Processing Systems}}}\ (\bibinfo {year} {2023})\BibitemShut {NoStop}%
\end{thebibliography}
%

\end{document}